\documentclass[aps,prb,longbibliography,twocolumn,superscriptaddress]{revtex4-2}
\usepackage{amsmath,amssymb,bm,graphicx,color,gensymb,bbold,hyperref,keyval,url,latexsym}
\usepackage[dvipsnames,svgnames,table]{xcolor}
\usepackage{enumitem}

\usepackage{hyperref}
\hypersetup{
    pdfstartview={FitH},
    colorlinks=true,
    linkcolor=NavyBlue,
    citecolor=Maroon,
    filecolor=NavyBlue,
    urlcolor=NavyBlue
}

\begin{document}

% ==============================================================================
%\title{Minimally-augmented spin-wave theory: application to the honeycomb-lattice models}
\title{Demystifying quantum escapism on the honeycomb lattice}
% ==============================================================================
\author{A. L. Chernyshev}
\affiliation{Department of Physics and Astronomy, University of California, Irvine, California 92697, USA}
% ==============================================================================
\date{\today}
% ==============================================================================
\begin{abstract}
We demonstrate the versatility, simplicity, and power of the minimally-augmented spin-wave theory in studying phase diagrams of the quantum spin models in which unexpected magnetically ordered phases occur or the existing ones expand beyond their classical stability regions. We use this method to obtain approximate phase diagrams of the two paradigmatic spin-$\frac{1}{2}$ models on the honeycomb lattice: the $J_1$--$J_3$ ferro-antiferromagnetic and $J_1$--$J_2$ antiferromagnetic $XXZ$ models. For the $J_1$--$J_3$ case, various combinations of the $XXZ$ anisotropies are analyzed. In a dramatic deviation from their classical phase diagrams, which host significant regions of the noncollinear spiral phases, quantum fluctuations stabilize several unconventional collinear phases and significantly extend conventional ones to completely supersede spiral states. These results are in close agreement with the available density-matrix renormalization group calculations. The applicability of this approach to the other models and its potential extension to different types of orders are discussed.
\end{abstract}
% ==============================================================================
\maketitle
%-------------------------------------------------------------------------------
\section{Introduction}
%-------------------------------------------------------------------------------

Frustrated spin systems are a cradle of exotic quantum states. With spin liquids being the best known~\cite{balents_2010,Norman_16,Kanoda_17,Knolle_19,Savary_2016},  multipolar phases~\cite{nematic-our,nematic-mike,PencLauchli2011,LucileS_1,Judit_recent} and nonmagnetic valence-bond solids (VBS)~\cite{ReadSachdev1989a,ReadSachdev1990b,Becca_01,Becca_17,Doretto20} also attract significant attention. For the unusual magnetic orders, the order-by-disorder (ObD) mechanism~\cite{Henley_89,Rau_obd,Knolle_25}, which selects a unique ground state from the classically degenerate manifold by an entropic criterion~\cite{Shender1982,Henley_87,Chubukov1992,Henley1992,Holdsworth92,%
ReimersBerlinsky1993,Henley1994,George_Mike04,Zhitomirsky2012,Savary2012,McClarty2014,%
kagome14}, is also much discussed. 

Rather undeservedly, the phenomenon of quantum selection of a magnetically ordered but completely unexpected ground state has received less attention. It is responsible for a class of quantum states, whose existence is also insufficiently acknowledged---states that are not a part of an accidentally degenerate manifold, if there was one, and states that are unrelated to any obvious instabilities that can be anticipated from the surrounding phases in the phase diagram of the given model. 

Usually, a magnetically ordered phase in a quantum spin model is associated with its counterpart in the classical limit of the same model. The unexpected magnetic phases break this association as they occur without having such classical counterparts. It is this phenomenon which we would like to refer to as ``quantum escapism.'' 

One can think of extending the model by a term which favors the unexpected phase and makes it the ground state in the classical limit somewhere in the extended parameter space, making the occurrence of such a state less mystifying, at least in principle. The remaining mysteries are the often dramatic extension of such phases from their nominal regions of stability and  the reason of why some states proliferate more readily than the others. 

Somewhat puristically, {\it all} aforementioned exotic quantum states can also be seen as the escapist states that are extending from some model extensions~\cite{DonnaSheng,Wietek}. The expansion of the magnetization plateau~\cite{Starykh_review} from a single classical point and the ObD selection of a state can also be viewed as a proliferation of the favored state from an extended model where it is a natural ground state.

Needless to say, this consideration also connects quantum escapism of the unexpected phases to the less exotic and more familiar expansion of the ordered quantum phases beyond their classical boundaries.
 
While this perspective is useful, the {\it method} to construct quantum phase diagrams, in which phases expand beyond their initially defined boundaries, is an open problem. More specifically, for  spin models, the problem of how to describe {\it magnetically ordered} quantum states beyond their classical regions of stability does not have a general solution. If achieved even approximately, such a description could yield quantitative insights into the ground state phase diagrams for a variety of models.  

In this work, we promote the utility of an approximate, physically well-justified, technically simple, and numerically inexpensive method that addresses this problem. The method was originally proposed and applied to the stabilization of quantum states in the transverse-field Ising model~\cite{Mila_12,Mila_14} and to the field-induced plateaus in the triangular and square lattices~\cite{Mila_13}. It is coined minimally-augmented spin-wave theory (MAGSWT), as it extends the standard SWT beyond the classical stability limits by introducing a minimal shift of the magnon chemical potential to stabilize it. 

Here, we demonstrate that MAGSWT can be successfully applied to a wide variety of magnetically ordered states in quantum spin models, yielding approximate phase diagrams of the two representative spin-$\frac{1}{2}$ honeycomb-lattice models: $J_1$--$J_3$ ferro-antiferromagnetic (FM-AF) and $J_1$--$J_2$ antiferromagnetic (AF) $XXZ$ models, in which several unexpected magnetically ordered phases appear and the existing ones expand beyond their classical stability regions. Some of the results for the $J_1$--$J_3$ model presented in this work were briefly reported in Ref.~\cite{j1j3} in conjunction with the density-matrix renormalization group (DMRG) study.

We choose to focus on these models to demonstrate the power of MAGSWT for several reasons. Both models have been  thought as harboring spin-liquid phases in their phase diagrams due to the low coordination number of the honeycomb lattice and strong frustration~\cite{Oitmaa91,Oitmaa92,j1j2j3-ed2001,Cabra11,Oitmaa11,%
j1j2-gong,j1j2-rigol1,j1j2-rigol2,j1j2-sheng,Arun23,Trebst2022}. Their actual description paints a significantly more complex picture. The classical $J_1$--$J_2$--$J_3$ model was known to host a variety of spiral states, also forming a classically degenerate manifold of them in the $J_1$--$J_2$ case~\cite{Rastelli79}. The ObD effect in this manifold was discussed more recently~\cite{j1j2-arun}, but numerical studies uncovered unexpected magnetic and VBS phases instead~\cite{j1j2-steve1,j1j2-steve2,j1j2-bishop}. 

The honeycomb-lattice spin systems have also attracted considerable attention in the search for Kitaev magnets~\cite{Witczak_Krempa14,Winter_17,Khaliullin18,Khaliullin20,Ross18,Park22,Lynn23,Armitage23}. However, many material realizations appear to be closely described by  a simpler $J_1$--$J_3$  $XXZ$ model with ``mixed''  (FM-AF) couplings, motivating its recent studies~\cite{Arun23,Trebst2022,LP90,BACAO}. 

One of the most outstanding examples of an unexpected escapist quantum state, which is also exceedingly unnatural, is the Ising-z (Iz) phase, first discovered numerically in the $XY$ limit of the $J_1$--$J_2$  honeycomb-lattice model~\cite{j1j2-steve2}. It was also recently found numerically in the $J_1$--$J_3$ FM-AF model~\cite{j1j3}. In this state, the ordered moments point along the $z$ axis despite the model having no out-of-plane $S^zS^z$  interactions in the $XY$ limit. Although magnetically ordered, the Iz state avoids breaking the $U(1)$ symmetry of the model, and has no classical counterpart within that model. It is also notable that the Iz state is unrelated to any state from the classically degenerate manifold of the co-planar spirals of the $J_1$--$J_2$ model, or the non-degenerate spiral state in the $J_1$--$J_3$ model, whose regions of their classical phase diagrams it occupies, nor is it anticipated by any magnon instability in the semi-classical analysis of these models. 

A rationalization of such an escapist state has pointed to a strong frustration in the $x$-$y$ plane and potentially large quantum fluctuations that lower the energy of the Iz state below that of the competing ones~\cite{j1j2-steve2}, making it escape-worthy. A fermionic description of this state was proposed~\cite{Glazman,Sedrakyan22}, suggesting a coexistence of the out-of-plane spin ordering with a chiral spin liquid. However, the rationalization provided above has never been supported quantitatively from the most natural perspective of the magnetically ordered state. In this work, we offer explicit demonstration of the large contribution of quantum fluctuations to the energy of the Iz state, which make it competitive for the ground states in both models.

Lastly, one of the important constraints on the use of the MAGSWT method is that the state to be stabilized should be an extremum, such as a saddle point. In practice, this translates to the absence of linear bosonic terms in the $1/S$-expansion as a sufficient criterion for the applicability of MAGSWT. In the absence of the bond-dependent Kitaev-like terms and Dzyaloshinskii-Moriya (DM) interactions, the collinearity of the state is a sufficient condition for the use of MAGSWT. As was found in the DMRG studies~\cite{j1j3,j1j2-steve1,j1j2-steve2}, all  unexpected escapist magnetic phases in the $J_1$--$J_2$ and $J_1$--$J_3$ models are collinear, making our analysis of the magnetic phase diagrams of the chosen models complete.

Our main results show that quantum fluctuations radically alter the classical phase diagrams for both models. In the $J_1$--$J_3$ model, two unexpected collinear phases, double-zigzag (dZZ) and Iz, are stabilized between the FM and zigzag (ZZ) phases, which also extend well beyond their classical regions, and the noncollinear spiral phase is completely eliminated in the $S\!=\!\frac{1}{2}$ limit. In the $J_1$--$J_2$ model, the classically degenerate spiral region is also eliminated in favor of a combination of N\'eel, stripe (collinear AF), and Iz phases, with the intermediate VBS phases not accessible by our approach but known from DMRG and other studies~\cite{j1j2-steve1,j1j2-arun}. We find that the MAGSWT phase boundaries closely track those obtained from state-of-the-art DMRG calculations, where the latter are available, demonstrating that this analytical method can reliably identify the correct ground-state order and even quantitatively estimate transition points. 
 
This method also provides significant quantitative insight into the energetics of the quantum stabilization of the non-classical phases, the competition between various states, and the role of the fluctuation contribution to their energies, also offering a systematic path for the explorations of similar models. Another demystifying aspect of this work is the systematic elimination of the noncollinear states, such as spirals, which are less effective at benefiting from quantum fluctuations, in favor of the collinear ones. This trend is in broad agreement with the arguments of the ObD phenomena~\cite{Henley_89,Rau_obd}, which generally favor collinear phases.

The rest of the paper is organized as follows. In Sec.~\ref{Sec:MagSWT}, we outline the MAGSWT method and its theoretical justification. In Sec.~\ref{Sec:J1J3}, we apply MAGSWT to the $J_1$--$J_3$ model: we describe the model and its classical phase diagram,  present the quantum phase diagram, and discuss how each phase is stabilized by fluctuations. Section~\ref{Sec:J1J2} addresses the $J_1$--$J_2$ model, highlighting the role of classical degeneracies and the resulting quantum phase selection. Finally, Sec.~\ref{Sec:Conclusions} summarizes our findings and suggests future directions, including possible extensions of the method to more complex states and other systems. 

%-------------------------------------------------------------------------------
\section{MAGSWT}
\label{Sec:MagSWT}
%-------------------------------------------------------------------------------

The energy minimization of a classical spin model provides ranges of the model parameters in which different states achieve an absolute energy minimum, yielding the classical phase diagram. The SWT approach consists of taking advantage of these classical ground states to develop a systematic $1/S$-expansion using a bosonic representation of spin operators~\cite{hp1940}, in which the local direction of the classical spin serves as a quantization axis. 

Since the classical energy is at a minimum, the terms that are linear in bosonic operators are guaranteed to vanish, the first non-zero term of the expansion is quadratic (harmonic), and the higher-order terms constitute various forms of interaction between bosonic spin excitations~\cite{RMP_13}. The purpose of this procedure is to study spin excitations, find quantum contributions to the ground-state energies, and take into account various other quantum effects within the ordered magnetic phases---the tasks at which such a spin-wave theory (SWT) is usually highly successful, both qualitatively and quantitatively. 

However, one such task is the description of the shift of the magnetic phase boundaries due to quantum fluctuations, which is also related to the quantum stabilization of the unexpected phases, as discussed above. In this regard, the standard SWT fails quite miserably already at the harmonic level of the $1/S$-expansion, because it requires calculating quantum contributions to the states  outside their classical regions of stability.  

To address this issue, the strategies to ``correct'' the unstable state by including higher-order $1/S$-terms, with or without the selfconsistency~\cite{Takahashi89,Chubukov_1991,Ivanov92,Gochev94,Gochev95,Alicea09,Takano_2011,%
us_CoNb2O6,MZh_fcc_22}, have been employed. Such calculations are tedious, have to be developed on a case-by-case basis, and their selfconsistency is rarely achieved outside the domain of the high-symmetry models and gapped states. 

A different, much simpler resolution of this general conundrum, which has plagued the application of the SWT to the classically unstable states, was originally suggested in Refs.~\cite{Mila_12,Mila_13,Mila_14} and is the basis of the present work. It is simple, elegant, and well-justified. We outline it below.

%-------------------------------------------------------------------------------
\subsection{Problem}
\label{Sec:problem}
%-------------------------------------------------------------------------------

Generally, for a stable classical minimum, the quadratic bosonic Hamiltonian in the SWT approach can be written in momentum space as 
\begin{equation}
\label{eq_H_SWT}
{\cal H}=E_{cl}+\frac{1}{2}\sum_{{\bf q}} \Big(\hat{\bf x}_{\bf q}^\dagger 
\hat{\bf H}_{\bf q}\hat{\bf x}_{\bf q}^{\phantom{\dagger}}
-\frac{1}{2}{\rm tr}(\hat{\bf H}_{\bf q})\Big)+O(S^0),
\end{equation}
where $E_{cl}$ is the classical energy, $O(S^2)$, $\hat{\bf x}^\dag_{\bf q}\!=\!\big( \hat{\bf a}^\dag_{\bf q}, 
\hat{\bf a}^{\phantom \dag}_{-{\bf q}}\big)$ is a vector of the bosonic creation and annihilation operators of length $2 n_s$, with $n_s$ being the number of bosonic species associated with the sublattices of the magnetic unit cell, $\hat{\bf H}_{\bf q}$ is a $2 n_s\!\times 2 n_s$ Hamiltonian matrix, $O(S)$, and ${\bf q}$ in the magnetic Brillouin zone. In this basis,
\begin{eqnarray}
\label{eq_LSWTmatrix}
\hat{\bf H}_{\bf q}=
\left( \begin{array}{cc} 
\hat{\bf A}^{\phantom \dagger}_{\bf q} &  \hat{\bf B}^{\phantom \dagger}_{\bf q}\\[0.5ex] 
\hat{\bf B}^\dagger_{\bf q}  & \hat{\bf A}^*_{\bf -q}
\end{array}\right),
\end{eqnarray}
where $\hat{\bf A}^{\phantom \dagger}_{\bf q}$ and $\hat{\bf B}^{\phantom \dagger}_{\bf q}$ 
are the $n_s \!\times n_s$ blocks of $\hat{\bf H}_{\bf q}$ corresponding to $\hat{\bf a}^\dag_{\bf q}\hat{\bf a}_{\bf q}$ and $\hat{\bf a}^\dag_{\bf q}\hat{\bf a}^\dag_{-\bf q}$ terms, respectively~\cite{Colpa,TothLake}. The diagonalization of $\hat{\bf g} \hat{\bf H}_{\bf q}$, where $\hat{\bf g}$ is the diagonal paraunitary $2 n_s\!\times 2 n_s$ matrix $\hat{\bf g}\!=\![\hat{\bf I},-\hat{\bf I}]$, with $\hat{\bf I}$ being the $n_s\!\times n_s$ identity matrix, yields $2 n_s$ linear SWT (LSWT) magnon eigenenergies 
$\{\varepsilon_{1,{\bf q}},  \varepsilon_{2,{\bf q}}, \dots,  -\varepsilon_{1,-{\bf q}},  -\varepsilon_{2,-{\bf q}},
\dots\}$~\cite{Colpa,TothLake}.

From (\ref{eq_H_SWT}), the energy of the ground state, to the order $O(S)$, is given by 
\begin{equation}
\label{eq_E}
{\cal E}\!=\!E_{cl}+\delta E,
\end{equation}
where $\delta E$ is the $1/S$ quantum contribution to the ground-state energy, with $\nu\!=\!1\dots n_s$,
\begin{equation}
\label{eq_dE}
\delta E=\frac{1}{2}\sum_{\bf q} \Big(\sum_{\nu}\varepsilon_{\nu,{\bf q}}
-{\rm tr}(\hat{\bf A}_{\bf q})\Big).
\end{equation}
As parameters of the model are varied, the classical state may cease to be a minimum, and the quadratic Hamiltonian in (\ref{eq_H_SWT}) stops being positive definite, as some of the $\varepsilon_{\nu,{\bf q}}^2$ of the matrix $(\hat{\bf g} \hat{\bf H}_{\bf q})^2$ become negative for some regions of the momenta ${\bf q}$. In fact, the search for the boundaries between classical phases can often be done by looking at such instabilities in the SWT spectra instead of the classical energy minimization~\cite{PRX_anisotropic_tri_19}. Needless to say, the $1/S$ quantum  contribution in (\ref{eq_dE}) becomes ill-defined outside the classical region of stability of the state. 

The root of the problem is clear:  the $1/S$-expansion is built upon a stable classical state, and if the latter ceases to be a ground state, i.e., becomes unstable, the LSWT eigenenergies $\varepsilon_{\nu,{\bf q}}$ are not well-defined. 

%%-------------------------------------------------------------------------------------------
\begin{figure}[t]
\includegraphics[width=\linewidth]{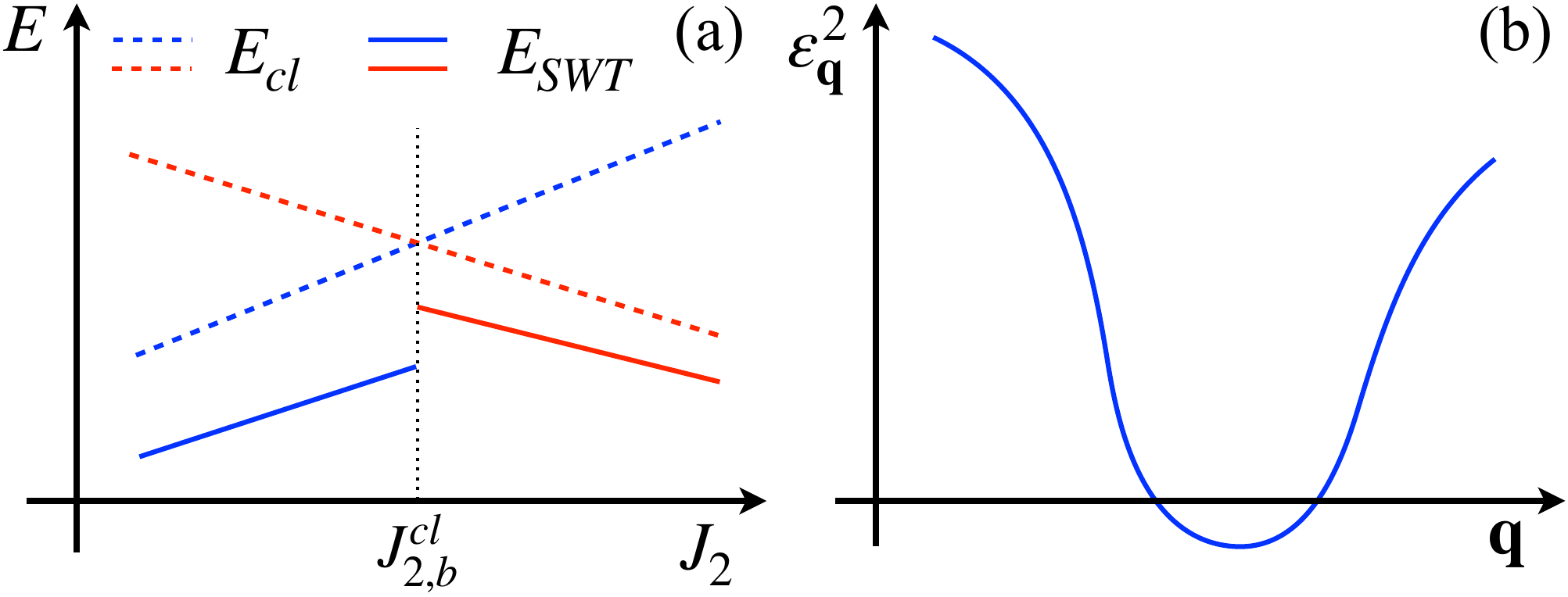}
\vskip -0.2cm
\caption{(a) Schematic illustration of the problem of the phase boundary within the standard SWT. The dashed and solid lines are classical and order $O(S)$ energies, respectively, dotted line marks classical phase boundary $J_{2,b}^{cl}$. (b) Schematics of $\varepsilon_{\nu,{\bf q}}^2$ calculated beyond the classical stability region.}
\label{fig:sketch}
\vskip -0.4cm
\end{figure}
%%-------------------------------------------------------------------------------------------

Figure~\ref{fig:sketch}(a) provides a qualitative illustration of the problem of the shift of the phase boundary due to quantum fluctuations within the standard SWT. The classical energies of the two ground states (dashed lines)  vs hypothetical model parameter $J_2$ cross at $J_{2,b}^{cl}$, which is the classical phase boundary. Generally, the energies of these states acquire different quantum contributions (\ref{eq_dE}), resulting in the  energies shown by solid lines in Fig.~\ref{fig:sketch}(a). While, clearly, the energy crossing should shift to a larger $J_{2}$, the calculation of $\delta E$ beyond the classical stability region is problematic for either of the states, because some of the $\varepsilon_{\nu,{\bf q}}^2$ become negative; see Fig.~\ref{fig:sketch}(b) for a sketch.

%-------------------------------------------------------------------------------
\subsection{MAGSWT resolution}
\label{Sec:resolution}
%-------------------------------------------------------------------------------

The resolution of this problem~\cite{Mila_12,Mila_13,Mila_14} consists of adding a local-field term to the Hamiltonian in the form
\begin{equation}
\label{eq_dH}
\delta{\cal H}=\mu\sum_{i} \left(S-{\bf S}_i\cdot{\bf n}_i\right),
\end{equation}
where ${\bf n}_i$ is the direction of the ordered moment in the classical spin configuration, which is also the local spin-quantization axis. In the bosonic language, this term is simply a shift of the chemical potential 
\begin{equation}
\label{eq_dHa}
\delta{\cal H}=\mu\sum_{i} a^\dag_i a_i.
\end{equation}
This form ensures that the classical energy in the expansion (\ref{eq_H_SWT}) is not altered and that $\mu$  in (\ref{eq_dHa}) only provides an additive constant to the diagonal elements of the Hamiltonian matrix $\hat{\bf H}_{\bf q}$ in (\ref{eq_LSWTmatrix}). Then, the {\it minimal positive} value of $\mu$ is found from the condition that all  eigenvalues $\varepsilon_{\nu,{\bf q}}^2$ of the matrix $(\hat{\bf g} \hat{\bf H}_{\bf q})^2$ are positive definite for all the momenta ${\bf q}$.

Once such a minimal $\mu$ is found, the energy ${\cal E}$ of the proposed spin state, Eq.~(\ref{eq_E}), with the $1/S$ contribution from Eq.~(\ref{eq_dE}), is  well-defined and can be compared with the energies of the competing states, which are calculated to the same $O(S)$  order. That is the essence of the {\it minimally-augmented spin-wave theory} (MAGSWT). While we will provide more technical details into the ways of finding the minimal shifts of the chemical potential for various states in the next Sections, let us first go through the list of benefits, strong aspects, limitations, and concerns about this approach.

The power of the method is not only in its simplicity, but in the form of the local-field term in Eq.~(\ref{eq_dH}), which guarantees that its contribution to the energy is positive definite for $\mu\!\geq\!0$. In turn, this implies that the so-obtained ground-state energy ${\cal E}$ is an {\it upper bound} for the true energy of such a state. In other words, if there is an exact solution for a given ground-state energy, which is expanded in $1/S$  to $O(S)$ order, that energy will necessarily be lower than the one obtained by MAGSWT. 

In the original works, Refs.~\cite{Mila_12,Mila_13,Mila_14}, this method was described as variational, which is not quite correct as the determination of the minimal $\mu$ does not involve any explicit minimization. However, given the statements of the MAGSWT energy being an upper bound, one can perceive it as variational in a generalized sense.  

One may be concerned that the shift of the chemical potential can ``prop up'' a state, while such a state would {\it not} have had a chance of becoming a true ground state otherwise. Let us dispel this concern.

\begin{figure}[t]
\includegraphics[width=\linewidth]{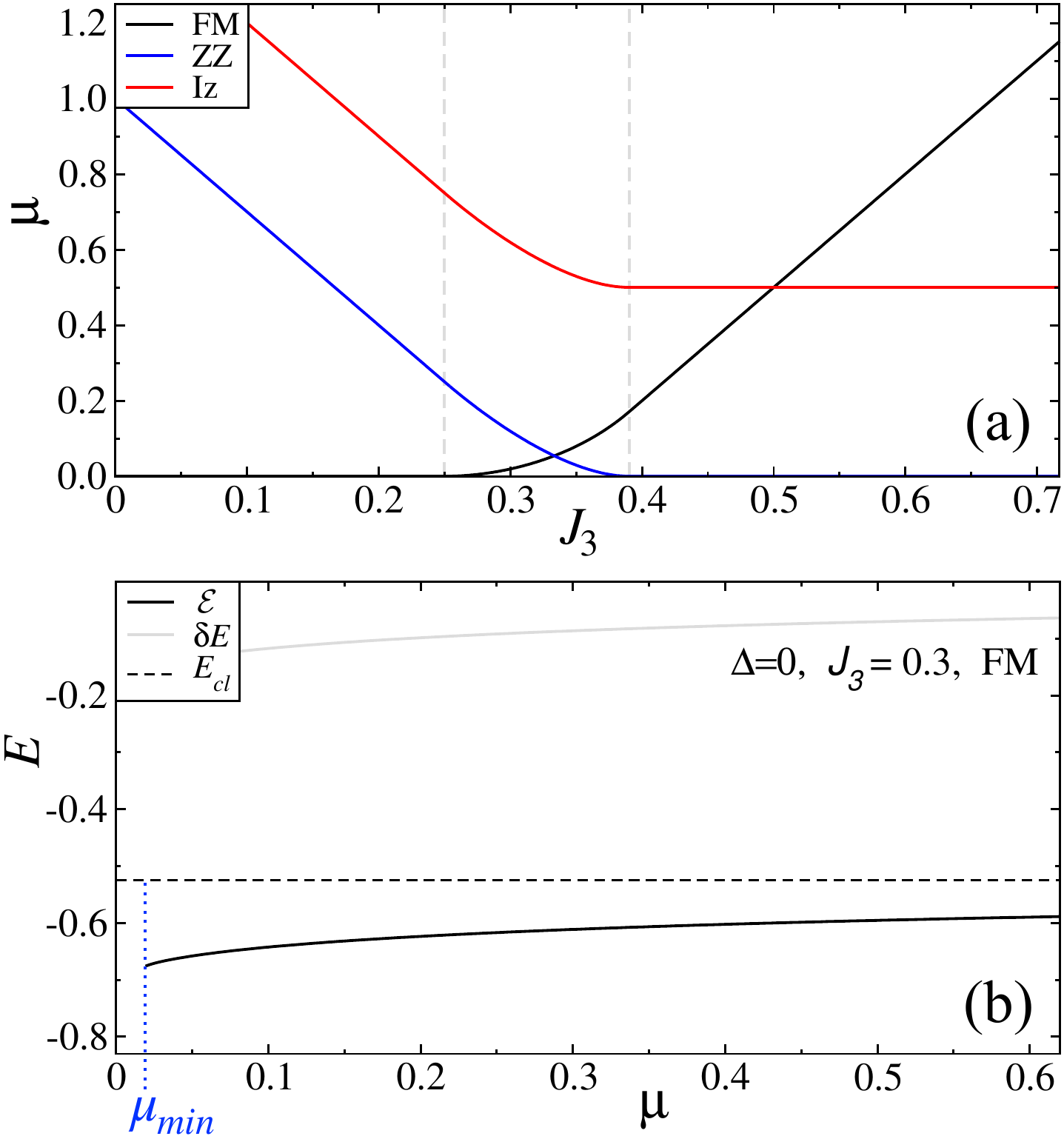}
\vskip -0.2cm
\caption{(a) The minimal $\mu$ for each of the three different states, FM, ZZ, and Iz, in the $J_1$--$J_3$ model for $\Delta\!=\!0$ and  $\Delta_3\!=\!1$ as a function of $J_3$, see Sec.~\ref{Sec:J1J3} for details. (b) The energy of the FM state for the same model at $J_3\!=\!0.3$ as a function of $\mu$. The minimal $\mu_{min}$ from (a) is indicated. Dashed line is the classical energy, solid lines are the quantum contribution $\delta E$, Eq.~(\ref{eq_dE}),  and the total energy ${\cal E}$, Eq.~(\ref{eq_E}), respectively.}
\label{fig:mu_vs_J3}
\vskip -0.4cm
\end{figure}

Omitting details that will be discussed in Sec.~\ref{Sec:J1J3}, in Fig.~\ref{fig:mu_vs_J3}(a) we show the minimal $\mu$ for three different states as a function of the model parameter ($J_3$ in this case). The two phases, FM and ZZ, are stable for $J_3\!\leq\!0.25$ and $J_3\!\agt\!0.39$, respectively, so their corresponding $\mu$ is zero in these regions, but is monotonically increasing away from their boundaries. The Iz phase is not classically stable anywhere, so its $\mu$ is non-zero throughout this 1D phase diagram.  These are the typical results.

In Fig.~\ref{fig:mu_vs_J3}(b), we plot the energy of the FM state, calculated by MAGSWT, as a function of $\mu$ from the region where the FM state is not stable classically, $J_3\!>\!0.25$. The calculations of the quantum contribution (\ref{eq_dE}) are physical only for $\mu\!\geq\!\mu_{min}$, as is explained above. One can see that $\delta E$ and the total energy (solid lines) are monotonic functions of $\mu$. It is easy to show that for $\mu\!\rightarrow\!\infty$, the quantum contribution in Eq.~(\ref{eq_dE}) approaches zero from below as $O(1/\mu)$, and the energy of the stabilized state simply reaches its classical value. Since the state that needs to be stabilized is not a minimum in the classical limit, it is obvious that it cannot be stabilized if it requires a  ``push'' with large $\mu$. 

One can also consider small-$\mu$ limit for a classically stable phase. A simple algebra in Eq.~(\ref{eq_dE}) or directly in Eq.~(\ref{eq_dHa}) gives ${\cal E}_{\rm MAGSWT}-{\cal E}_{\rm SWT}\!=\!\mu\cdot \delta \lambda \!>\!0$, where $\delta \lambda$ is the reduction of the ordered moment by quantum fluctuations and we have assumed that the ordered moment is the same on all sites. 

Combined with the argument provided above that the MAGSWT energies should serve as the upper bound to the true energies of a state to the order $O(S)$, this discussion helps to demonstrate that MAGSWT cannot prop up an arbitrary state to become the ground state if such a state has no potential to be one, $\mu$ notwithstanding. Conversely, if quantum fluctuations can stabilize the state, MAGSWT provides a reasonable estimate of its energy in the stabilized regime. In this way, the method allows one to explore candidate phases beyond their classical stability limits and to assess which phase might become the ground state as parameters change.

%%-------------------------------------------------------------------------------------------
%%-------------------------------------------------------------------------------------------

We note that MAGSWT in its present form cannot be applied to an arbitrary state outside the classical region of its stability; a necessary criterion is that the classical spin state is an extremum of the energy, such as a saddle point~\cite{Mila_13}, so that no linear bosonic terms appear in its spin-wave expansion. In practice, this restricts MAGSWT to collinear or high-symmetry noncollinear states and models without the bond-dependent Kitaev-like or DM terms. 

In our models, all phases of interest are collinear and satisfy this criterion. The  spiral states considered in this work interpolate continuously between the collinear ones. Therefore, they are fully confined to their classical regions of existence and do not require MAGSWT to be considered on the same $O(S)$ footing with the other phases since the standard SWT suffices.

The advantages of MAGSWT are several-fold. It allows one to extend calculations of the quantum-corrected ground-state energies of various states beyond their classical regions of stability, unlike the standard SWT approach.  It provides a reasonably straightforward and computationally inexpensive way to construct the phase diagrams of the quantum models and analyze phases that are unexpected from the classical considerations or from instabilities of the spin-wave spectra of the neighboring phases. The phase diagram and the phase boundaries are determined from the comparison of the energies ${\cal E}$ and finding their intersections for all the considered competing phases as a function of the varied model parameters. MAGSWT also provides physical insight by essentially quantifying how much the energy of each candidate phase can be lowered due to fluctuations. 

In particular, MAGSWT is naturally consistent with the idea that quantum fluctuations preferentially stabilize collinear states, in line with the ObD arguments~\cite{Henley_89,Rau_obd}. It provides reasonable explanations of the often dramatic expansions of the collinear phases and of why they do so more readily than other phases.

In the following Sections, we will apply MAGSWT to specific models and demonstrate quantitatively its ability to map out the quantum phase diagrams and provide insights into the energetics of the competing states.

%-------------------------------------------------------------------------------
\section{$J_1$--$J_3$ FM-AF Model}
\label{Sec:J1J3}
%-------------------------------------------------------------------------------

Some of the results for the MAGSWT energies and phase diagrams for this model were briefly reported previously in Ref.~\cite{j1j3} and its Supplemental Material. Here, we provide actual technical details and discussions that are essential for the practical use of the method, and extend the parameter space for the $J_3$ coupling.  

%-------------------------------------------------------------------------------
\subsection{Model and some background}
\label{Sec:J1J3model_history}
%-------------------------------------------------------------------------------

Interestingly, some of the earliest studies of the mixed FM-AF $J_1$--$J_2$--$J_3$ honeycomb-lattice models, which date back to the 1970s, Ref.~\cite{Rastelli79}, were motivated by some of the same materials~\cite{LP90} that have received significant renewed interest today in the context of the search for the Kitaev magnets~\cite{BACAO}.

There is also a close similarity between the {\it classical} phase diagram of this model and that of the pure AF  $J_1$--$J_2$--$J_3$ model on the same lattice, which has been the focus of the searches for exotic quantum states more recently, but still in the pre-Kitaev era~\cite{Oitmaa91,Oitmaa92,j1j2j3-ed2001,j1j2-arun,Oitmaa11,Cabra11,j1j2-rigol2,j1j2-rigol1,j1j2-bishop,j1j2-gong,j1j2-steve1,j1j2-steve2}. That search was motivated by the expectation of  stronger fluctuations due to the lattice's low  coordination number  and  by the degeneracies in its  classical phase diagram~\cite{j1j2-arun}. With some of these degeneracies discussed below in Sec.~\ref{Sec:J1J2}, we also note that Ref.~\cite{Rastelli79} has observed that the spin-wave spectrum  in the spiral phases may contain low-energy branches, indicating some near-degeneracies of the ground state already at the level of the quasiclassical consideration,  but the fluctuation corrections to the classical ground-state energies of different phases have not been closely discussed until more recently~\cite{j1j2-arun}.  

In the surge of interest in honeycomb-lattice magnets over the last decade, despite the possible presence of the Kitaev-like terms in their microscopic models, it appears that the minimal $XXZ$-anisotropic $J_1$--$J_3$ model with mixed FM-AF exchanges provides a  close description for many of these materials~\cite{Rethinking,LP90,usLPR,arun-firstprinciple,pavel-dZZ,Broholm_2023,Trebst2022,Arun23}, calling for a deeper study of this model.

The anisotropic $XXZ$  $J_1$--$J_3$ FM-AF model on the honeycomb lattice is given by
\begin{eqnarray}
\label{eq_H}
{\cal H}=\sum_{n=1,3}\sum_{\langle ij\rangle_n} J_n  \Big(S^{x}_i S^{x}_j+S^{y}_i S^{y}_j+\Delta_n S^{z}_i S^{z}_j\Big),
\end{eqnarray}
where the nearest-neighbor FM exchange, $J_1\!=\!-1$, is used as the energy unit,  the third-nearest-neighbor $J_3$ is AF, $J_3\!>\!0$, and $\langle ij\rangle_n$ denotes $n$th-neighbor bonds. The anisotropic $XXZ$ version of the model (\ref{eq_H}), which is of most interest, corresponds to the easy-plane regime, $0\!\leq\!\Delta_{1(3)}\!\leq\!1$, with the $x$ and $y$ axes forming the spins' easy-plane and $z$ axis is orthogonal to it. 

The standard choice is to make anisotropies the same in the $J_1$ and $J_3$  terms, $\Delta_{1}\!=\!\Delta_{3}$, which will be referred to as the {\it full $XXZ$} model. However, because in real materials  further exchanges tend to be more isotropic, we will also focus on a different version of the $XXZ$ model, with $J_3$ in the Heisenberg limit, $\Delta_3\!=\!1$, referred to as the {\it partial $XXZ$} model. As one will see, the phase diagram is somewhat richer in this case. Obviously, these two models coincide in the Heisenberg limit of $\Delta\!=\!1$. In addition, we will also interpolate the $XY$ limits of these models by considering $\Delta_1\!=\!0$ for the  FM term and varying $\Delta_{3}$ from 0 to 1, from the $XY$ to the Heisenberg limit. 

%-------------------------------------------------------------------------------
\subsection{Classical phases and phase diagram}
\label{Sec:J1J3classical}
%-------------------------------------------------------------------------------

\begin{figure}[t]
\includegraphics[width=\linewidth]{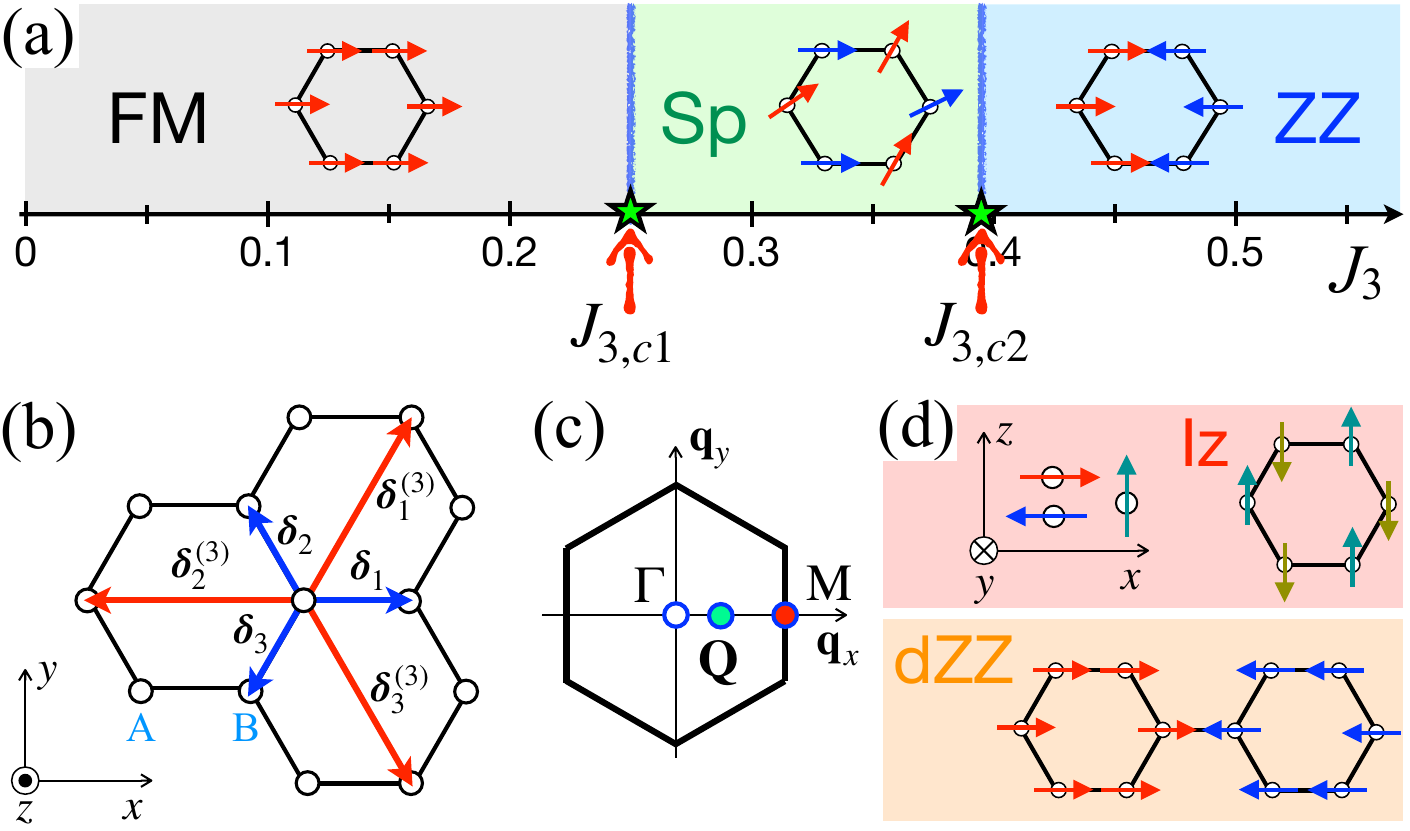}
%\vskip -0.2cm
\caption{(a) Classical phase diagram of the $J_1$--$J_3$ model (\ref{eq_H}) for any $0\!\leq\!\Delta_{1(3)}\!\leq\!1$. Sketches of the FM, Sp, and ZZ illustrate the spin order in each phase (red and blue arrows belong to two sublattices). The transition points $J_{3,c1}\!=\!0.25$ and $J_{3,c2}\!\approx\! 0.3904$ are indicated. (b) The sketch of the honeycomb lattice lattice with A and B sublattices, crystallographic axes, and the nearest- and third-neighbor vectors, ${\bm \delta}_\alpha$ and ${\bm \delta}^{(3)}_\alpha$. (c) Brillouin zone (BZ) of the honeycomb lattice with the high-symmetry $\Gamma$ and M points and the representative ${\bf Q}$ vector of the spiral. (d) Sketches of the Iz and dZZ states. Axes in the upper panel show out-of-plane spin direction in the Iz phase and in-plane for the other phases.}
\label{Fig_1D}
%\vskip -0.3cm
\end{figure}

The classical phase diagram and LSWT excitation spectra of the classical phases of the model (\ref{eq_H}) were first studied in Ref.~\cite{Rastelli79}. Since all states minimizing classical energy are coplanar with the $x$-$y$ plane, the classical phase diagram shown in Fig.~\ref{Fig_1D}(a) is the same for any choice of the $XXZ$ anisotropy in either of the terms, $0\!\leq\!\Delta_{1(3)}\!\leq\!1$.  It consists of three phases: ferromagnetic (FM) and zigzag (ZZ) orders in the regimes dominated by $J_1$ and $J_3$, respectively, interpolated by a planar co-rotating spin spiral (Sp), in which spins in the two sublattices of the honeycomb lattice are offset by a finite angle; see below for more detail. Classically, both the FM-Sp transition at $J_{3,c1}\!=\!0.25$ and Sp-ZZ transition at $J_{3,c2}\!=\!(\sqrt{17}-1)/8$ are continuous, meaning that the Sp state continuously interpolates between the FM and ZZ states, having them as limiting cases of the spiral~\cite{Rastelli79}.

% ==============================================================================
% ==============================================================================
 
For the dominant FM $J_1$, the FM state is obvious. For large $J_3$, the choice of the ZZ state, which consists of the ferromagnetic zigzag chains arranged antiferromagnetically,  is also intuitive because the AF $J_3$ couples different sublattices of the original honeycomb lattice; see Fig.~\ref{Fig_1D}(b). For $J_1\!=\!0$, the network of $J_3$ couplings forms three independent N\'eel-ordered honeycomb lattices, with a finite FM $J_1$ ordering them in the ZZ fashion.

%which also shows the sketch of the lattice, A and B sublattices, crystallographic axes, and the nearest- and third-neighbor vectors, ${\bm \delta}_\alpha$ and ${\bm \delta}^{(3)}_\alpha$. 

The appearance of the Sp state as a classical ground state is less obvious, and can be obtained either from the energy minimization or from the instabilities of the magnon spectra in the FM and ZZ phases. Such instabilities in both cases correspond to softening of the Goldstone mode at the ordering vectors of these phases at $J_{3,c1}$ and $J_{3,c2}$, respectively, and both FM and ZZ magnon branches becoming complex for some regions of ${\bf q}$-space in between these $J_3$ values.  

A simple algebra yields the classical energies of  the FM and ZZ states within the model (\ref{eq_H}),  per number of atomic unit cells of the honeycomb lattice $N_A$ and in units of $|J_1|$, as given by
\begin{align}
\label{eq_Ecl}
E_{cl}^{\rm FM}=- S^2(3-3J_3),\ \ \ E_{cl}^{\rm ZZ}=-S^2(1+3J_3).
\end{align}
Note that these expressions are, indeed, independent of $\Delta_{1(3)}$, and are valid for any $J_3$, inside or outside the states' stability regions. 

Using the single-${\bf Q}$ ansatz for the Sp state, with some algebra, one can reproduce the result  of Ref.~\cite{Rastelli79}. The spiral state is a single-${\bf Q}$ state with the ordering vector ${\bf Q}\!=\!(Q_x,0)$ with spins in the two sublattices offset from each other by an angle $\varphi$
\begin{eqnarray}
\label{eq_Qx}
&&Q_x=\frac{2}{3}\cdot\arccos\left(\frac{1}{2J_3}\cdot\frac{1-3J_3}{1-2J_3}\right),\\
&&\varphi=-\frac{Q_x}{2}+\arctan\left(\frac{(1-J_3)\sin (3Q_x/2)}{2+(1-3J_3)\cos (3Q_x/2)}\right), \nonumber
\end{eqnarray}
with all momenta in units of $1/a$, the inverse nearest-neighbor lattice distance of the honeycomb lattice. One can verify that the ordering vector ${\bf Q}$ in (\ref{eq_Qx}) is continuously migrating as a function of $J_3$ from the FM ordering vector ${\bf Q}\!=\!\Gamma\!=\!(0,0)$ at $J_{3,c1}\!=\!0.25$ to that of the ZZ phase ${\bf Q}\!=\!{\rm M}\!=\!(2\pi/3,0)$ at $J_{3,c2}\!=\!(\sqrt{17}-1)/8\!\approx\!0.3904$, in agreement with the discussion above; see Fig.~\ref{Fig_1D}(c), which shows  Brillouin zone (BZ) of the honeycomb lattice with the high-symmetry $\Gamma$ and M points~\cite{Rastelli79}. 

The classical energy of the spiral state,  per number of atomic unit cells of the honeycomb lattice $N_A$ and in units of $|J_1|$, is 
\begin{align}
\label{eq_Ecl_sp}
E_{cl}^{\rm Sp}=-3S^2\left(\Re\left[e^{i\varphi}\gamma_{-\bf Q}\right]
-J_3\Re\left[e^{i\varphi}\gamma^{(3)}_{-\bf Q}\right]\right),
\end{align}
with  ${\bf Q}$ and $\varphi$ defined in (\ref{eq_Qx}) and the nearest- and third-neighbor hopping amplitudes given by
\begin{align}
\label{eq_gks}
\gamma_\mathbf{q}&=\frac{1}{3}\sum_{\alpha} e^{i\mathbf{q}{\bm \delta}_\alpha},\ \ \ 
\gamma^{(3)}_{\mathbf{q}}=\frac{1}{3}\sum_{\alpha} e^{i\mathbf{q}{\bm \delta}^{(3)}_\alpha},
\end{align}
with the primitive vectors ${\bm \delta}_\alpha$ and ${\bm \delta}^{(3)}_\alpha$ shown in Fig.~\ref{Fig_1D}(b).

As is demonstrated by the  DMRG results in Ref.~\cite{j1j3} for the  $S\!=\!\frac12$  $J_1$--$J_3$ model (\ref{eq_H}), this classical picture is incomplete in the quantum case, as some unexpected collinear phases are stabilized in its phase diagram. In addition, the existing collinear phases, FM and ZZ,  also extend beyond their classical ranges of stability, while the noncollinear Sp state is absent altogether. 

Therefore, for the purpose of the subsequent MAGSWT treatment, we need to consider two other states, which are not the classical ground states,  in addition to the FM, ZZ, and Sp.  These are the double-zigzag (dZZ) and Ising-$z$ (Iz) states, see Fig.~\ref{Fig_1D}(d). In the former, {\it two} subsequent zigzag columns of the ferromagnetically aligned spins order antiferomagnetically. In the latter,  spins escape the  frustrated coupling within the $x$-$y$ plane and order antiferomagnetically in the standard N\'eel fashion along the out-of-plane $z$ axis, leaving the U(1) symmetry of the $XXZ$ model intact. This phase was first discovered in the $XY$ $J_1$--$J_2$ AF model on the same lattice~\cite{j1j2-steve2}, which will be discussed in Sec.~\ref{Sec:J1J2}.

% ==============================================================================
\begin{figure}[t]
\includegraphics[width=\linewidth]{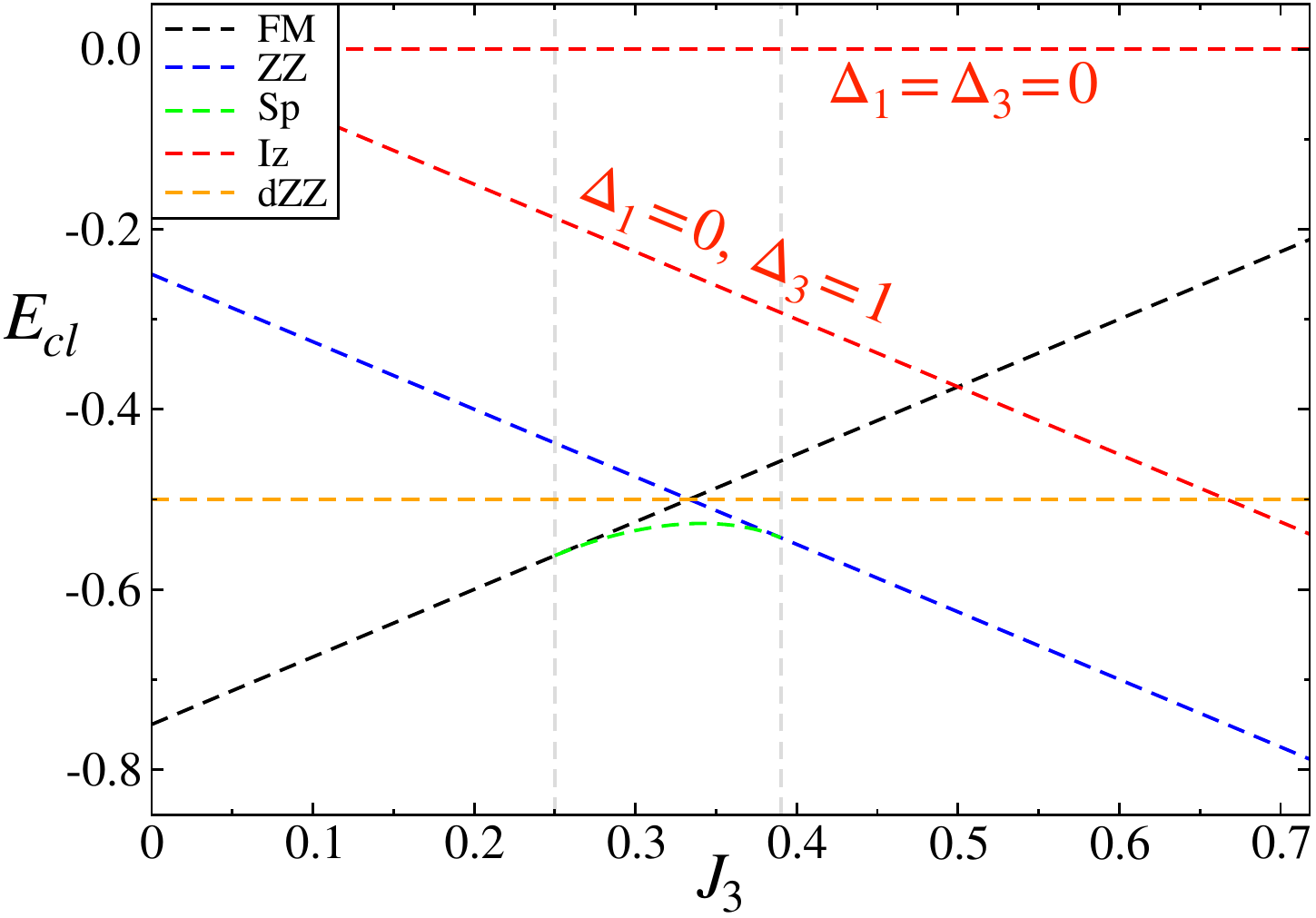}
\vskip -0.2cm
\caption{Classical energies of the FM, ZZ, Sp, Iz, and dZZ states as a function of $J_3$ from Eqs.~(\ref{eq_Ecl}), (\ref{eq_Ecl_sp}), and (\ref{eq_Ecl_add}) for $S\!=\!\frac12$. Vertical dashed lines are the FM-Sp and Sp-ZZ transitions, see Fig.~\ref{Fig_1D}(a). For the Iz state, two limiting cases are shown, $\Delta_1\!=\!\Delta_3\!=\!0$ and $\Delta_1\!=\!0$, $\Delta_3\!=\!1$.}
\label{Fig_Ecl}
\vskip -0.3cm
\end{figure}
% ==============================================================================

The classical energies of these two phases are
\begin{align}
\label{eq_Ecl_add}
E_{cl}^{\rm Iz}=S^2(3\Delta_1-3J_3\Delta_3),\ \ \ E_{cl}^{\rm dZZ}=-2S^2,
\end{align}
per number of atomic unit cells of the honeycomb lattice $N_A$ and in units of $|J_1|$, as before. 

In Figure~\ref{Fig_Ecl}, we show classical energies of all five states from Eqs.~(\ref{eq_Ecl}), (\ref{eq_Ecl_sp}), and (\ref{eq_Ecl_add}), in the relevant range of $J_3$ and using spin $S\!=\!\frac12$. Vertical dashed lines mark the FM-Sp and Sp-ZZ transitions, and the shown results for all states except for Iz are independent of the $XXZ$ anisotropies within the discussed easy-plane regime. The Sp state is continuously interpolating FM and ZZ states, as discussed above. For the Iz state, we show the energies in the two limiting cases: both $J_1$ and $J_3$ terms are $XY$ ($\Delta_1\!=\!\Delta_3\!=\!0$) and the ``partial'' $XY$ limit with $J_3$ term isotropic ($\Delta_1\!=\!0$, $\Delta_3\!=\!1$). The latter case provides the lowest boundary for the classical energy of the Iz state within the model (\ref{eq_H}).

One can see that even in the best case of the ``partial'' $XY$ limit the Iz state is considerably higher than the rest of the states. On the other hand, the dZZ is low enough to be an important suspect for the quantum escapism. Moreover, if not for the Sp state, which takes over the intermediate $J_3$ region, one can see that the crossing point of the FM and ZZ energies at $J_3\!=\!1/3$ is also the crossing with the dZZ state. In fact, there is a hidden classical degeneracy at this point: all states comprised of the FM zigzag chains that are arranged either ferromagnetically or antiferromagnetically, are degenerate at  $J_3\!=\!1/3$. 

Although in the classical model this degeneracy point is intercepted by the spiral phase, in the quantum limit the Sp phase is suppressed and such a close degeneracy seems to be surfacing up in the form of the close competition between different states, as is evidenced by the long 1D-like ferromagnetic correlations observed in DMRG; see Refs.~\cite{j1j3,Arun23}. This effect may also be relevant to the phenomenology of BaCo$_2$(AsO$_4$)$_2$, a material receiving significant recent interest~\cite{Broholm_2023,BACAO}. In its case, a small magnetic field, associated with very small energy scale, is capable of switching the dZZ state into a mix of ZZ and dZZ states, with the up-up-down alternating directions of the FM zigzag chains~\cite{LP90,Regnault_18}. 

%-------------------------------------------------------------------------------
\subsection{LSWT}
\label{Sec:J1J3SWT}
%-------------------------------------------------------------------------------

Here we elaborate on the standard LSWT details for all  classical states in order to pave the way for the MAGSWT,  discussed next.

Of the five states shown in Fig.~\ref{Fig_1D}(a) and \ref{Fig_1D}(d) and discussed above, the unit cell of the magnetic structure for the FM and Iz states is that of the atomic unit cell of the honeycomb lattice, containing two sites. For the ZZ and Sp states, the unit cell can be reduced to the atomic one using the staggered or rotated reference frames, respectively. For the dZZ state, the staggered reference frame reduces the magnetic unit cell from eight to four sites. Thus, the Hamiltonian matrix $\hat{\bf H}_{\bf q}$ in Eq.~(\ref{eq_LSWTmatrix}) is $4\times 4$  in the first four cases, while for the dZZ state it is  $8\times 8$.

Since a very similar $4\!\times 4$  matrix structure appears in the LSWT treatment of the phases of interest in the $J_1$--$J_2$ model discussed in Sec.~\ref{Sec:J1J2}, here we recall the general expressions for the Hamiltonian's  eigenenergies for it. 

In all two-sublattice cases considered in this work, the LSWT matrices $\hat{\bf A}_{\bf q}$, $\hat{\bf B}_{\bf q}$ in Eq.~(\ref{eq_LSWTmatrix})  assume the same form 
\begin{eqnarray}
\label{eq_AB_2sub}
\hat{\bf A}^{\phantom \dag}_{\bf q}=
\left( \begin{array}{cc} 
A_{\bf q}  &  B_{\bf q}\\
B^*_{\bf q}  & A_{\bf q}
\end{array}\right), \quad \quad 
\hat{\bf B}^{\phantom \dag}_{\bf q}=
\left( \begin{array}{cc} 
D_{\bf q} &  C_{\bf q}\\
C^*_{\bf q}  & D_{\bf q}
\end{array}\right),
\end{eqnarray}
with the matrix elements $A_{\bf q}$ and $D_{\bf q}$ being purely real, $B_{-\bf q}\!=\!B^*_{\bf q}$, and $C_{-\bf q}\!=\!C^*_{\bf q}$. In this case, the eigenvalues of the Hamiltonian matrix~\eqref{eq_LSWTmatrix} can be found analytically by diagonalizing $\big(\hat{\bf g}\hat{\bf H}_{\bf q}\big)^2$ instead of $\hat{\bf g}\hat{\bf H}_{\bf q}$, giving two magnon branches residing in the full BZ~\cite{j1j2-arun,Rastelli79}
\begin{align}
\varepsilon_{\nu,{\bf q}}&=\sqrt{A_{\bf q}^2-D_{\bf q}^2+|B_{\bf q}|^2-|C_{\bf q}|^2 +(-1)^\nu R}\ ,\nonumber\\
\label{eq_E12_2sub}
&R=2\sqrt{|A_{\bf q}B_{\bf q}-C_{\bf q}D_{\bf q}|^2-[\Im(B_{\bf q} C^*_{\bf q})]^2}\ .
\end{align}
For the two-sublattice cases, FM, ZZ, Iz, and Sp, considered here, there are additional simplifications, such as $D_{\bf q}\!=\!0$ and $A_{\bf q}\!=\!A$. For the FM and Iz states and in all four cases  in the  limit $\Delta_{1(3)}\!=\!0$, one can also find additional simplifications of the eigenvalue problem using transformations that reduce the  $4\times 4$  matrix to the block-diagonal form of $2\times 2$  matrices~\cite{kopietz,maksimov16,maksimov22},  but we mostly refrain from discussing these details. 

%-------------------------------------------------------------------------------
\subsubsection{FM}
\label{Sec:FM}
%-------------------------------------------------------------------------------

In the FM case,  the matrix  elements are given by
\begin{align}
A&=3S(1-J_3) , \ \ D_{\bf q}=0,\nonumber\\
B_{\bf q}&=-3S\big(\left(1+\Delta_1\right)\gamma_{\bf q} -J_3\left(1+\Delta_3\right)\gamma^{(3)}_{\bf q} \big)/2,\nonumber\\
\label{eq_ABC_FM}
C_{\bf q}&=-3S\big(\left(1-\Delta_1\right)\gamma_{\bf q} -J_3\left(1-\Delta_3\right)\gamma^{(3)}_{\bf q} \big)/2,
\end{align}
with the hopping amplitudes given in Eq.~(\ref{eq_gks}). 

Aside from some obvious simplification in the Heisenberg limit of both terms, in which the off-diagonal terms vanish altogether, there is another case that is  useful for the subsequent MAGSWT insights: the ``full'' $XY$ limit,  $\Delta_1\!=\!\Delta_3\!=\!0$. In this case, $B_{\bf q}\!=\!C_{\bf q}$, and the two branches are, explicitly
\begin{align}
\label{eq_E12_XYFM}
\varepsilon_{\nu,{\bf q}}&=3S\sqrt{(1-J_3)\big(1-J_3+(-1)^\nu\left|\bar{\gamma}_{\bf q}\right|\big)},
\end{align}
where $\bar{\gamma}_{\bf q}\!=\!\gamma_{\bf q}-J_3\gamma^{(3)}_{\bf q}$. It is clear that the second bracket in the lower magnon branch contains the potentially ``offending'' element, which is responsible for the softening of the spectrum at $J_{3,c1}\!=\!0.25$ and for the negative $\varepsilon_{1,{\bf q}}^2$ for $J_{3}\!>\!J_{3,c1}$. 

%-------------------------------------------------------------------------------
\subsubsection{ZZ}
\label{Sec:ZZ}
%-------------------------------------------------------------------------------

In the ZZ case,  the matrix  elements are 
\begin{align}
A&=S(1+3J_3) ,\ \ D_{\bf q}=0, \nonumber\\
B_{\bf q}&=-3S\big(\gamma_{\bf q}-\Delta_1\gamma'_{\bf q} -J_3\left(1-\Delta_3\right)\gamma^{(3)}_{\bf q} \big)/2,\nonumber\\
\label{eq_ABC_ZZ}
C_{\bf q}&=-3S\big(\gamma_{\bf q}+\Delta_1\gamma'_{\bf q} -J_3\left(1+\Delta_3\right) \gamma^{(3)}_{\bf q}\big)/2,
\end{align}
where $\gamma'_\mathbf{q}=\left( e^{i\mathbf{q}{\bm \delta}_1}-e^{i\mathbf{q}{\bm \delta}_2}-e^{i\mathbf{q}{\bm \delta}_3}\right)/3.$ 
As in the FM case, there are several simplifications possible, with the ``full'' $XY$ limit being similarly instructive
\begin{align}
\label{eq_E12_XYZZ}
\varepsilon_{\nu,{\bf q}}&=S\sqrt{(1+3J_3)\big(1+3J_3+3(-1)^\nu\left|\bar{\gamma}_{\bf q}\right|\big)},
\end{align}
containing the same $\bar{\gamma}_{\bf q}\!=\!\gamma_{\bf q}-J_3\gamma^{(3)}_{\bf q}$ element. 

%-------------------------------------------------------------------------------
\subsubsection{Iz}
\label{Sec:Iz1}
%-------------------------------------------------------------------------------

In the Iz case,  the matrix  elements are given by
\begin{align}
A&=-3S(\Delta_1-\Delta_3 J_3) ,\ \ \ B_{\bf q}=D_{\bf q}=0,\nonumber\\
\label{eq_ABC_Iz}
C_{\bf q}&=-3S\big(\gamma_{\bf q} -J_3 \gamma^{(3)}_{\bf q}\big)=-3S\bar{\gamma}_{\bf q},
\end{align}
yielding two degenerate branches
\begin{align}
\label{eq_E12_Iz}
\varepsilon_{1,{\bf q}}=\varepsilon_{2,{\bf q}}=\sqrt{A^2-\left|C_{\bf q}\right|^2},
\end{align}
with the same $\bar{\gamma}_{\bf q}$ element persistently present. 

%-------------------------------------------------------------------------------
\subsubsection{Sp}
\label{Sec:Sp1}
%-------------------------------------------------------------------------------

In the Sp case,  the matrix  elements are given by
\begin{align}
\label{eq_ABC_Sp}
A&=3S\left(\Re\left[e^{i\varphi}\gamma_{-\bf Q}\right]
-J_3\Re\left[e^{i\varphi}\gamma^{(3)}_{-\bf Q}\right]\right) ,\\
B_{\bf q}&=-\frac{3S}{2}\bigg(\Delta_1\gamma_{\bf q}+
\frac{1}{2}\big(e^{i\varphi}\gamma_{{\bf q}-\bf Q}+e^{-i\varphi}\gamma_{{\bf q}+\bf Q}\big)\nonumber\\
&\  \quad-J_3\Big[\Delta_3\gamma^{(3)}_{\bf q}+\frac{1}{2}\big(e^{i\varphi}\gamma^{(3)}_{{\bf q}-\bf Q}+e^{-i\varphi}\gamma^{(3)}_{{\bf q}+\bf Q}\big)\Big]\bigg),\nonumber\\
C_{\bf q}&=-\frac{3S}{2}\bigg(\Delta_1\gamma_{\bf q}-
\frac{1}{2}\big(e^{i\varphi}\gamma_{{\bf q}-\bf Q}+e^{-i\varphi}\gamma_{{\bf q}+\bf Q}\big)\nonumber\\
&\  \quad-J_3\Big[\Delta_3\gamma^{(3)}_{\bf q}-\frac{1}{2}\big(e^{i\varphi}\gamma^{(3)}_{{\bf q}-\bf Q}+e^{-i\varphi}\gamma^{(3)}_{{\bf q}+\bf Q}\big)\Big]\bigg),\nonumber
\end{align}
where $D_{\bf q}\!=\!0$ and we have generalized results of Ref.~\cite{Rastelli79}, which considered the limiting $XY$ and Heisenberg cases.

%-------------------------------------------------------------------------------
\subsubsection{dZZ}
\label{Sec:dZZ}
%-------------------------------------------------------------------------------

In the 4-sublattice dZZ case $\hat{\bf A}_{\bf q}$ and $\hat{\bf B}_{\bf q}$ matrices  are
\begin{eqnarray}
\label{eq_AB_4sub}
&&\hat{\bf A}^{\phantom \dag}_{\bf q}\!=\!
\left( \begin{array}{cccc} 
A_1  &  B_{\bf q} & 0  & D_{\bf q} \\
B^*_{\bf q}  & A_1 & D^*_{\bf q}  & 0\\
0 & D_{\bf q} & A_2 & C_{\bf q} \\
D^*_{\bf q} & 0 & C^*_{\bf q} & A_2
\end{array}\right), \\
&&\hat{\bf B}^{\phantom \dag}_{\bf q}\!=\!
\left( \begin{array}{cccc} 
0  &  C_{\bf q} & 0  & F_{\bf q} \\
C^*_{\bf q}  & 0 & F^*_{\bf q}  & 0\\
0 & F_{\bf q} & 0 & B_{\bf q} \\
F^*_{\bf q} & 0 & B^*_{\bf q} & 0
\end{array}\right), \nonumber
\end{eqnarray}
with the matrix elements given by 
\begin{align}
A_1&=S(3-J_3) , \ \ \ A_2=S(1+J_3) ,\nonumber\\
B_{\bf q}&=-\frac{S}{2}\Big((1+\Delta_1)\gamma_{1,\bf q} \nonumber\\
&\quad -J_3\big((1-\Delta_3)\gamma^{(3)}_{2,\bf q}+ (1+\Delta_3)\gamma^{(3)}_{13,\bf q}\big)\Big),\nonumber\\
C_{\bf q}&=-\frac{S}{2}\Big((1-\Delta_1)\gamma_{1,\bf q} \nonumber\\
&\quad -J_3\big((1+\Delta_3)\gamma^{(3)}_{2,\bf q}+ (1-\Delta_3)\gamma^{(3)}_{13,\bf q}\big)\Big),\nonumber\\
\label{eq_ABCDF_dZZ}
D_{\bf q}&=-\frac{S}{2}(1+\Delta_1)\gamma_{23,\bf q}, 
\ \ F_{\bf q}=-\frac{S}{2}(1-\Delta_1)\gamma_{23,\bf q},
\end{align}
where the hopping amplitudes are introduced as
\begin{align}
\gamma_{1,\bf q}&=e^{i{\bf q}{\bm \delta}_1},\ \
\gamma_{23,\bf q}=e^{i{\bf q}{\bm \delta}_2}+e^{i{\bf q}{\bm \delta}_3}, \nonumber\\
\label{eq_gks_dZZ}
\gamma^{(3)}_{2,\bf q}&=e^{i{\bf q}{\bm \delta}^{(3)}_2},\ \
\gamma^{(3)}_{13,\bf q}=e^{i{\bf q}{\bm \delta}^{(3)}_1}+e^{i{\bf q}{\bm \delta}^{(3)}_3}.
\end{align}
The eigenvalue problem for the $8\times 8$ $\big(\hat{\bf g}\hat{\bf H}_{\bf q}\big)^2$ matrix is not reducible to a compact analytical form in this case. However,  analytical solutions are available for the eigenenergies at the  high-symmetry ${\bf q}\!=\!0$ and  ${\bf q}\!=\!(0,\pi/\sqrt{3})$ points in the Heisenberg limit, which are instrumental for finding the MAGSWT parameter $\mu$, discussed next.

Needless to say, within the standard LSWT, the Iz and dZZ magnon energies  are imaginary in major parts of the BZ throughout the phase diagram, and so are the solutions for the FM and ZZ magnon branches outside their classical regions of stability. 

%-------------------------------------------------------------------------------
\subsection{Finding $\mu$}
\label{Sec:J1J3mu}
%-------------------------------------------------------------------------------

As is explained in Sec.~\ref{Sec:resolution}, the MAGSWT approach to stabilize the magnon spectra outside the classical boundaries consists of adding the {\it minimal} value shift of the bosonic chemical potential (\ref{eq_dHa}) that renders all magnon eigenvalues $\varepsilon^2_{\nu,{\bf q}}$ positive definite for all ${\bf q}$. Given the explicit expressions for the LSWT Hamiltonian matrices and eigenvalues for the specific states described above, the practical problem is to find  such a shift of the chemical potential as a function of the parameters of the model, and, preferably, in an analytical form. 

The Sp state interpolates between the FM and ZZ and corresponds to a minimum of the classical energy in its entire range of existence between the $J_{3,c1}$ and $J_{3,c2}$ bounds. As such, it is not the subject of the MAGSWT treatment, and its quantum energy contribution to Eq.~(\ref{eq_E}) is perfectly well-defined within the conventional LSWT. To be clear,  since Eq.~(\ref{eq_E}) yields the $O(S)$ energy for the Sp state, this energy can be faithfully compared to the $O(S)$ energies of all competing phases, obtained with or without the MAGSWT help. 

The other four states are either classically unstable altogether (Iz and dZZ), or need to be stabilized outside their classical ranges (FM and ZZ), so the MAGSWT needs to be used to calculate their energies. All four phases are collinear, which guarantees the absence of the linear bosonic terms in their $1/S$-expansion for the case of $XXZ$ interactions in the model (\ref{eq_H}).

We note that the limiting $XY$ and Heisenberg cases, as well as the examination of the magnon spectra at the select high-symmetry ${\bf q}$ points, appear very useful for obtaining  analytical expressions for  $\mu(J_3,\Delta_{1,3})$, eliminating the need of a numerical scan of the  momentum space for the spectrum instabilities. Once such a functional dependence of $\mu$ is found, the quantum contribution to the state's energy in Eq.~(\ref{eq_dE}) can be straightforwardly calculated, and the $O(S)$ energy surfaces ${\cal E}(J_3,\Delta_{1,3})$ in the model's parameter space can be readily obtained for each state. Then the MAGSWT phase boundaries are found from the intersections of such surfaces. 

%-------------------------------------------------------------------------------
\subsubsection{$\mu$ for FM, ZZ, and Iz}
\label{Sec:mu_FMZZIz}
%-------------------------------------------------------------------------------

For the FM, ZZ, and Iz states, the search for the minimal value of $\mu$ utilized a similar approach. In the limiting cases, such as ``full'' $XXZ$ ($\Delta_1\!=\!\Delta_3$), Heisenberg, and $XY$ limits, analytical expression for the magnon bands, such as the ones in Eqs.~(\ref{eq_E12_XYFM}), (\ref{eq_E12_XYZZ}), and (\ref{eq_E12_Iz}), simplify sufficiently to yield the explicit $J_3$-dependence of the offending negative minimum of the lowest branch $\varepsilon_{1{\bf q}}^2$ that needs to be lifted up by the positive shift in the regions where the spectrum is unstable. The required shift is easily related to $\mu$, while the $\Delta_{1,3}$-dependencies of $\mu$  appears to be either absent or to follow trivially from the considered limiting cases. 

In fact, in these three cases one finds that the condition for the lowest magnon mode to become stable can be expressed in a unified form. The diagonal elements $A$ of the LSWT matrices $\hat{\bf A}_{\bf q}$ before augmentation are given in Eqs.~(\ref{eq_ABC_FM}), (\ref{eq_ABC_ZZ}), and (\ref{eq_ABC_Iz}) for the FM, ZZ, and Iz state, respectively.  The resulting solutions  {\it for all three cases} correspond to a change $A\!\rightarrow\!\bar{A}$ with $\bar{A}$  being  
\begin{align}
\label{eq_mus}
\bar{A}=A+\mu=3S\left|\bar{\gamma}_{{\bf Q}_{\rm max}}\right|, 
\end{align}
where $\bar{\gamma}_{\bf q}\!=\!\gamma_{\bf q}-J_3\gamma^{(3)}_{\bf q}$, the structural element hinted upon in Sec.~\ref{Sec:J1J3SWT}, and ${\bf Q}_{\rm max}$ is the momentum at which $\left|\bar{\gamma}_{\bf q}\right|$ achieves maximal value for a given $J_3$. The condition for finding the maximum of $|\bar{\gamma}_{\bf q}|$ is equivalent to the search of the ordering vector associated with the classical energy minimum of the model (\ref{eq_H}).
Thus, unsurprisingly, ${\bf Q}_{\rm max}$ is equal to the ordering vectors of the FM phase for $J_3\!\leq\! J_{3,c1}$ and  of the ZZ phase for $J_3\!\geq\! J_{3,c2}$, while assuming the values of $(Q_x,0)$ given in Eq.~(\ref{eq_Qx}) for the $J_3$ range corresponding to the Sp phase.
 
Altogether,  ${\bf Q}_{\rm max}$ is defined piecewise as
\begin{align}
\label{eq_Qm}
{\bf Q}_{\rm max}&=
\left\{\begin{array}{ll} 
(0,0), &  \ \  J_3\leq J_{3,c1}, \\
(Q_x,0), & \ \  J_{3,c1}<J_3<J_{3,c2},\\
(2\pi/3,0), &  \ \ J_3\geq J_{3,c2}.
\end{array} \right.
\end{align}
Thus, for a given $J_3$, one can determine $\mu$ for the three states from Eqs.~(\ref{eq_mus}) and (\ref{eq_Qm}).

Technically, the definition of $\mu$ from Eq.~(\ref{eq_mus}) with ${\bf Q}_{\rm max}$ from (\ref{eq_Qm}) and (\ref{eq_Qx}) suffices. With some tedious, but straightforward algebra one can instead obtain compact expression for $\left|\bar{\gamma}_{{\bf Q}_{\rm max}}\right|$ explicitly in terms of $J_3$,
\begin{align}
\label{eq_bar_gamma_max}
\left|\bar{\gamma}_{{\bf Q}_{\rm max}}\right|&\!=\!
\left\{\begin{array}{ll} 
(1-J_3), &    J_3\!\leq\! J_{3,c1}, \\
\displaystyle{\frac13\sqrt{\frac{(1+2J_3)(1-J_3)^3}{J_3(1-2J_3)}}}, &  J_{3,c1}\!<\!J_3\!<\!J_{3,c2},\\
(1/3+J_3), &   J_3\!\geq\! J_{3,c2}.
\end{array} \right.
\end{align}
In Sec.~\ref{Sec:MagSWT}, we have provided a representative plot of the shift of the chemical potentials $\mu$ for the three phases discussed here; see Fig.~\ref{fig:mu_vs_J3}(a). We note that for the FM and ZZ states, both $\mu$ and the augmented diagonal matrix element $\bar{A}$ are independent of  the $XXZ$ anisotropies $\Delta_{n}$, with the solution~(\ref{eq_mus}) valid for any of them. 

Although the resulting $\mu$ for the Iz phase depends on the $XXZ$ parameters $\Delta_{1}$ and $\Delta_{3}$, it does so trivially, via the corresponding dependence of the LSWT matrix element $A$ in Eq.~(\ref{eq_ABC_Iz}), keeping the augmented $\bar{A}$ independent of them. Interestingly enough, the MAGSWT spectrum in the Iz case, and the quantum energy contribution (\ref{eq_E}) derived from it, are fully independent of the anisotropy parameters $\Delta_n$.

%-------------------------------------------------------------------------------
\subsubsection{$\mu$ for dZZ}
\label{Sec:mu_dZZ}
%-------------------------------------------------------------------------------

In the dZZ phase, solving for $\mu$ is more involved. We performed numerical diagonalization of the $8\times 8$ bosonic matrix $\big(\hat{\bf g}\hat{\bf H}_{\bf q}\big)^2$ matrix with $\hat{\bf A}_{\bf q}$ and $\hat{\bf B}_{\bf q}$ from (\ref{eq_AB_4sub}) for high-symmetry ${\bf q}\!=\!0$ and  ${\bf q}\!=\!(0,\pi/\sqrt{3})$ points in the Heisenberg limit to identify the soft modes as $J_3$ varies. The diagonalization at these points can be reduced to an analytical form that yields the minimal values of $\mu(J_3)$ in each region defined by those softening points. Altogether, the resultant explicit expressions for $\mu$ are 
\begin{align}
\label{eq_mu_dZZ}
\mu&=\!
\left\{\begin{array}{ll} 
S\left(\sqrt{5\!-\!2J_3\!+\!J_3^2}\!-\!1\!-\!3J_3 \right), &   J_3\!<\!\widetilde{J}_{c1}, \\
\mbox{interpolate}, & \widetilde{J}_{c1}\!<\!J_3\!<\!\widetilde{J}_{c2},\\
2S\left(\sqrt{2\!-\!2J_3\!+\!J_3^2} \!-\!1\right), &  \widetilde{J}_{c2}\!<\!J_3\!<\!\widetilde{J}_{c3},\\
2SJ_3, &  J_3\!>\!\widetilde{J}_{c3},
\end{array} \right.
\end{align}
with $\widetilde{J}_{c1}\!=\!0.1892$, $\widetilde{J}_{c2}\!=\!0.203$, and $\widetilde{J}_{c3}\!=\!0.25$. 
In a narrow interval $\widetilde{J}_{c1} \!<\! J_3 \!<\! \widetilde{J}_{c2}$, two lowest magnon branches alternately soften at small but finite ${\bf q}$'s, away from the high-symmetry points. For that region, we find that a linear interpolation for $\mu$ across that region is the most efficient, as it is sufficient to lift both instabilities, resulting in a nonzero but very small gap.

Following the other collinear phases, $\mu$ for the dZZ phase is independent of the $XXZ$ anisotropies, the property also verified numerically. 

With the MAGSWT strategy outlined in Sec.~\ref{Sec:MagSWT}, quantum contributions to the ground-state energies in all competing phases can now be calculated in a conventional $1/S$ fashion using Eq.~(\ref{eq_E}) with the  shifts of the chemical potential given in Eqs.~(\ref{eq_mus}) and (\ref{eq_mu_dZZ}). Then the $O(S)$ energies of the competing phases can be compared to create the phase diagram of the model (\ref{eq_H}). The results of this effort are provided next.

%-------------------------------------------------------------------------------
\subsection{Results, energies}
\label{Sec:J1J3resultsE}
%-------------------------------------------------------------------------------

With the results of Sec.~\ref{Sec:J1J3mu}, we can now calculate the $O(S)$ energies in Eq.~(\ref{eq_E}), ${\cal E}\!=\!E_{cl}+\delta E$, as a function of $J_3$ and anisotropies $\Delta_{1(3)}$ for all competing phases. Here we present some representative results illustrating such a competition along several $J_3$-cuts through the phase diagrams of the model (\ref{eq_H}) for different choices of the $XXZ$ parameters.  

\begin{figure}[t]
\includegraphics[width=\linewidth]{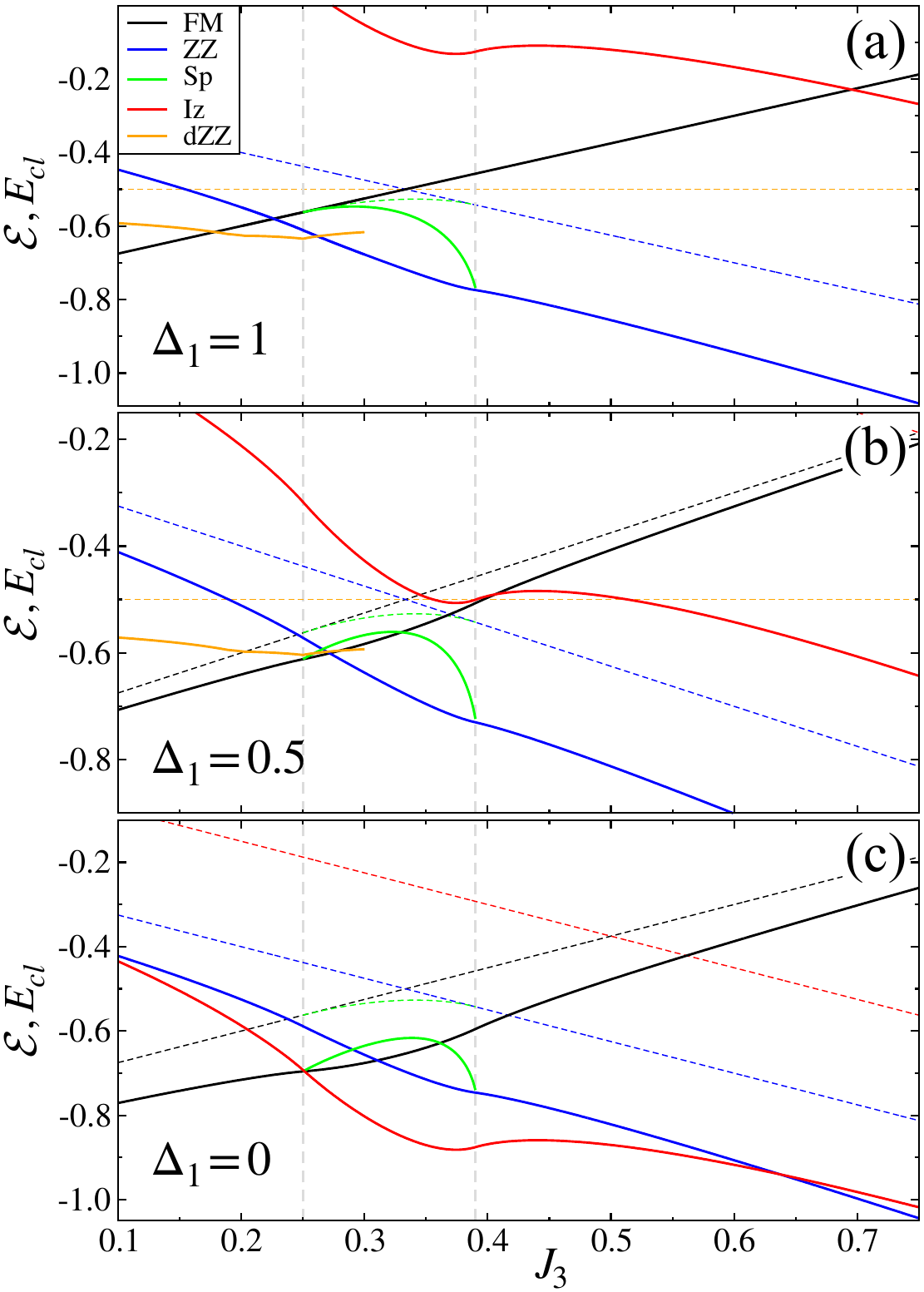}
\vskip -0.2cm
\caption{Energies of the FM, ZZ, Sp, Iz, and dZZ states as a function of $J_3$ for $\Delta_{3}\!=\!1$ and $S\!=\!1/2$. Dashed lines are the classical energies, Eqs.~(\ref{eq_Ecl}), (\ref{eq_Ecl_sp}), and (\ref{eq_Ecl_add}),  and solid lines are ${\cal E}\!=\!E_{cl}+\delta E$ from (\ref{eq_E}). Vertical dashed lines are classical transition boundaries from Fig.~\ref{Fig_1D}. (a) $\Delta_{1}\!=\!1$ (all Heisenberg limit), (b) $\Delta_{1}\!=\!0.5$, and (c)  $\Delta_{1}\!=\!0$ (partial $XY$ limit).} 
\label{Fig_Es_partial}
\vskip -0.3cm
\end{figure}

Figure~\ref{Fig_Es_partial} shows three $J_3$-cuts for the ``partial'' $XXZ$ case, in which $J_3$-term is kept  isotropic, $\Delta_{3}\!=\!1$. The dashed lines in all three panels are the  classical energies of the phases from Fig.~\ref{Fig_Ecl} and  Eqs.~(\ref{eq_Ecl}), (\ref{eq_Ecl_sp}), and (\ref{eq_Ecl_add}), and solid lines are the ${\cal E}$ energies obtained using Eq.~(\ref{eq_E}).  Vertical dashed lines mark the classical FM-Sp and Sp-ZZ boundaries, $J_{3,c1}$ and $J_{3,c2}$, from Fig.~\ref{Fig_1D}(a). 

In Fig.~\ref{Fig_Es_partial}(a), i.e., in the Heisenberg limit of the model (\ref{eq_H}), the FM state is an exact eigenstate, so the classical energy in (\ref{eq_Ecl}) is also exact and quantum corrections to it are zero, whether within its region of stability or not (black line). For all other states and for the FM state in Figs.~\ref{Fig_Es_partial}(b) and \ref{Fig_Es_partial}(c), quantum contributions $\delta E$ from Eq.~(\ref{eq_dE}) play essential role in lowering their energies. 

It is especially true for the Iz state in all three panels, with its classical energy lines falling outside the limits of Figs.~\ref{Fig_Es_partial}(a) and \ref{Fig_Es_partial}(b), and the downward renormalization of its energy being about two to three times larger than for any other  competing state. Indeed, $\delta E$ constitutes major term in the Is state's energy balance. In Figs.~\ref{Fig_Es_partial}(a)-(c), one can observe the steady progress of it toward becoming a ground state in a significant range of $J_3$ in the partial $XY$ limit of the model (\ref{eq_H}). 

% ==============================================================================
% ==============================================================================

For the dZZ state, the $O(S)$ energies from Eq.~(\ref{eq_E}) with the shift of the chemical potential from  Eq.~(\ref{eq_mu_dZZ}) still need to be obtained by a numerically more costly procedure than for the rest of the states, because they require diagonalization of the $8\times 8$ Hamiltonian matrix in Eq.~(\ref{eq_AB_4sub}). For that reason, the results for ${\cal E}$ for dZZ state in Fig.~\ref{Fig_Es_partial} are presented for a limited range of $J_3$ and only for $\Delta_1$ values where dZZ state is competitive. 

The trend for dZZ is opposite to that of the Iz state. Its  quantum contribution to energy is rather modest, and it can only compete for the ground state in the region of $\Delta_n$ close to the Heisenberg limit, where it can carve some of the $J_3$ range from the FM phase, which is  nearly free from quantum effects approaching this limit. In that carving of the fluctuation-free FM phase space, it also competes with the ZZ state, which fluctuates significantly.

Lastly, the Sp state is completely superseded by the neighboring ZZ and FM states in all  panels of Fig.~\ref{Fig_Es_partial},  showing that it is not as effective in lowering its energy as the competing states. Since it is  coincident with the FM and ZZ states at its limits, it is degenerate with them at these limiting points, but otherwise it is not benefiting enough from quantum fluctuations. This is in accord with the ObD  expectations that collinear states tend to be favored by fluctuations~\cite{Henley_89}.

Such a detailed  analysis of the energetics demonstrate  the ability of the MAGSWT approach to provide quantitative insights into the competition of the classically stable and unstable states on equal footing.

% ==============================================================================
\begin{figure}[t]
\includegraphics[width=\linewidth]{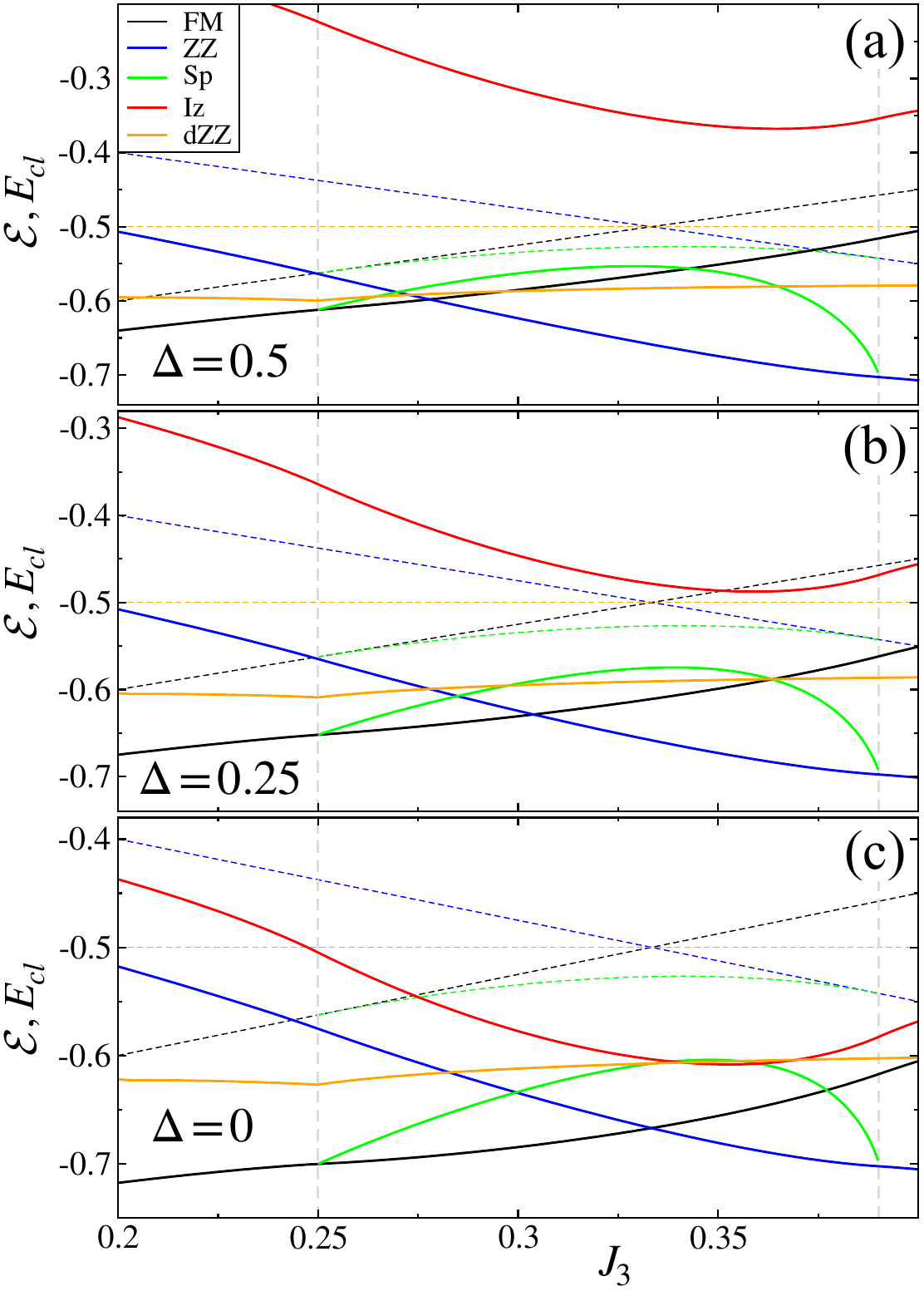}
\vskip -0.2cm
\caption{Same as in Fig.~\ref{Fig_Es_partial} for the FM, ZZ, Sp, Iz, and dZZ states in the ``full'' $XXZ$ model, $\Delta_{1}\!=\!\Delta_{3}\!=\!\Delta$, and in the narrower range of $J_3$. (a) $\Delta\!=\!0.5$, (b) $\Delta\!=\!0.25$, and (c)  $\Delta\!=\!0$ (full $XY$ limit).} 
\label{Fig_Es_full}
\vskip -0.3cm
\end{figure}
% ==============================================================================

Figure~\ref{Fig_Es_full} presents additional analysis of such kind for the  ``full'' $XXZ$ version of the model (\ref{eq_H}), $\Delta_{1}\!=\!\Delta_{3}\!=\!\Delta$, focusing on the narrower $J_3$ range where competition is the closest and on the $XXZ$ anisotropies toward the $XY$ limit of the model, $\Delta\!\leq\!0.5$. 

% ==============================================================================
\begin{figure}[t]
\includegraphics[width=\linewidth]{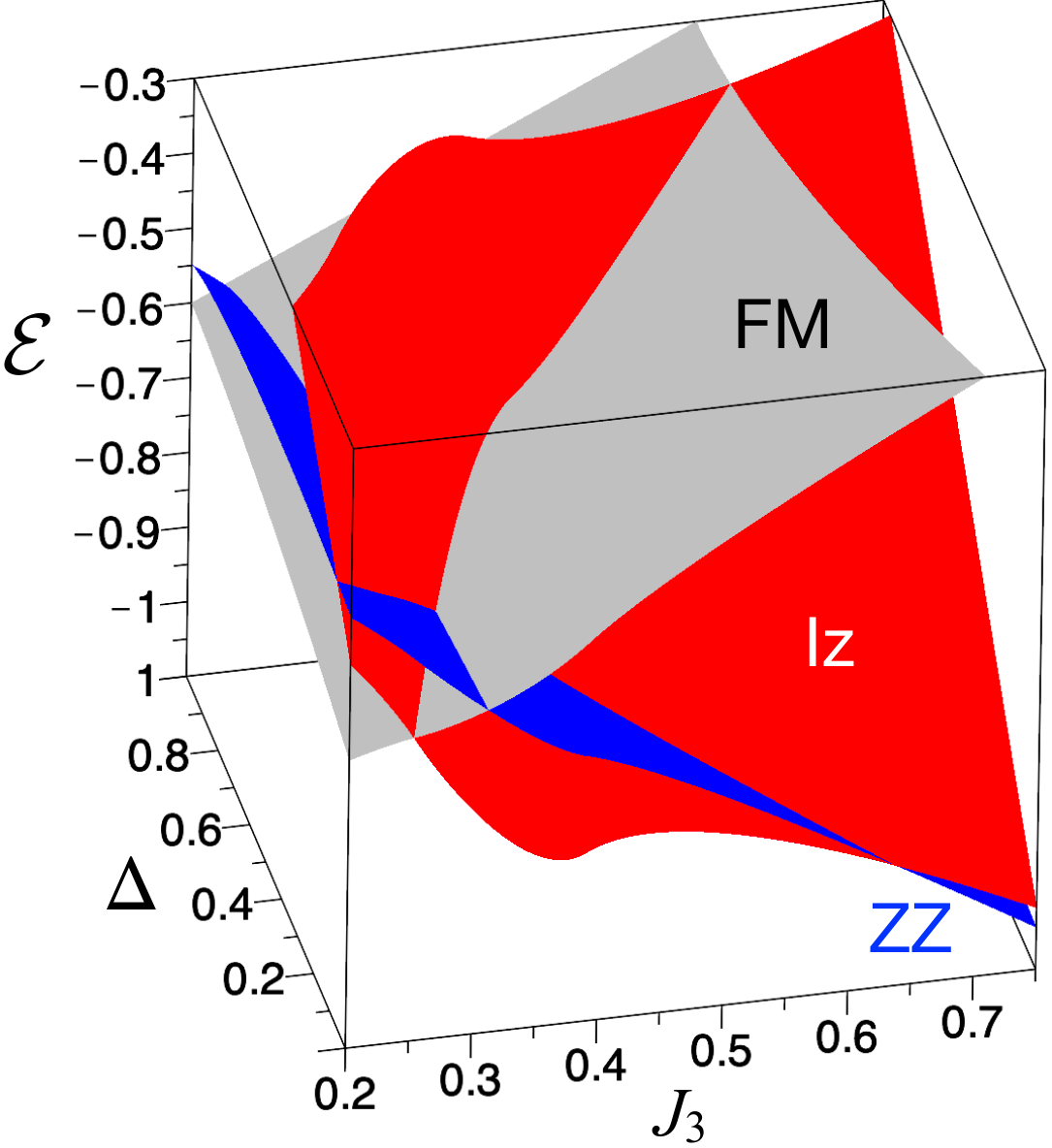}
\vskip -0.2cm
\caption{MAGSWT 2D energy surfaces ${\cal E}(J_3,\Delta_1)$,  Eq.~(\ref{eq_E}), for the FM, ZZ, and Iz states for the partial-$XXZ$ version of the model (\ref{eq_H}), $\Delta_3\!=\!1$,   for $S\!=\!1/2$.}
\vskip -0.3cm
\label{Fig_Es_3D}
\end{figure}
% ==============================================================================

While the trends are similar to the ones discussed for Fig.~\ref{Fig_Es_partial}, the resultant ground states in all three panels are FM and ZZ. The Sp state is not competitive for the same reasons as before. Fig.~\ref{Fig_Es_full}(a) corresponds to $\Delta\!=\!0.5$, which is close to the tip of the dZZ phase boundary. It is clear that despite the even stronger energy fluctuation contributions than in Fig.~\ref{Fig_Es_partial}, the Iz state in Fig.~\ref{Fig_Es_full}(c) is not able to reach the ground state without the isotropic component of the $J_3$ term. The latter is present in the ``partial'' $XXZ$ case of Fig.~\ref{Fig_Es_partial}(c),  providing an additional coupling to the out-of-plane spin components that helps Iz state in becoming the true ground state.

In the presented analysis of the stabilization of the classically unstable states, the dZZ state is shown to be an opportunistic one, benefiting from a close proximity of the classical degeneracy point between the FM and multi-zigzag states. It has been recently shown~\cite{BACAO} that anisotropic Kitaev-like terms expand the range of stability of this unexpected dZZ state further into  extended parameter space of such a model. 

In contrast, the Iz state is a daring escapist, attempting to win over the rest of the competing states by the shear force of quantum fluctuations, and succeeding significantly with a little help from the out-of-plane coupling. It is yet to be found in any real material. 

Once the analytical expressions for the shifts of the chemical potential $\mu(J_3,\Delta_n)$, Eqs.~(\ref{eq_mus}) and (\ref{eq_mu_dZZ}),  are found, the computational ease of finding the $O(S)$ energies by MAGSWT in the full parameter space of the phase diagram of the model (\ref{eq_H}) is rather remarkable. 

To demonstrate this feature explicitly, in Figure~\ref{Fig_Es_3D} we provide an example of the 2D energy surfaces ${\cal E}(J_3,\Delta_1)$ for the FM, ZZ, and Iz states for the partial-$XXZ$ version of the model (\ref{eq_H}), $\Delta_3\!=\!1$,   for $S\!=\!1/2$ case. From the intersects of such energy surfaces, we construct the  groundstate phase diagrams, presented next.

%-------------------------------------------------------------------------------
\subsection{Results, phase diagrams}
\label{Sec:J1J3resultsPhD}
%-------------------------------------------------------------------------------

\begin{figure}[t]
\includegraphics[width=\linewidth]{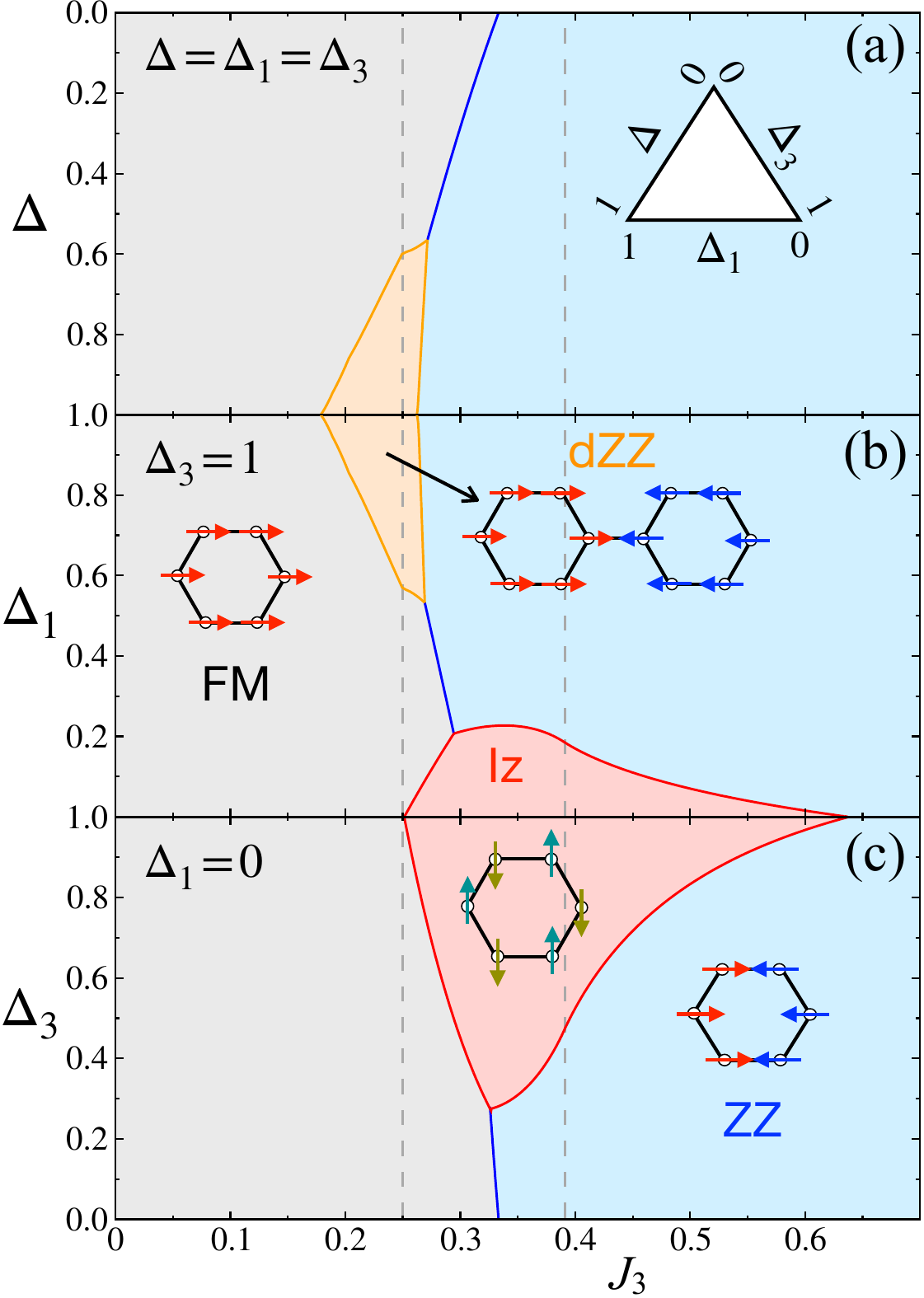}
\vskip -0.2cm
\caption{MAGSWT phase diagram for the (a) full $XXZ$, (b)  partial $XXZ$, and (c) $\Delta_1\!=\!0$ versions of the model (\ref{eq_H}) for $S\!=\!1/2$, with phases and their sketches identified. Vertical dashed lines are classical phase boundaries from Fig.~\ref{Fig_1D}. The inset in (a) shows the $\Delta_n$ path of the combined phase diagram.}
\label{Fig_PhDJ1J3}
\vskip -0.3cm
\end{figure}

We conclude this Section by the phase diagrams for different $XXZ$ versions of the model (\ref{eq_H}) shown in Figure~\ref{Fig_PhDJ1J3}. The phase boundaries are drawn from the  intersection lines of the calculated ${\cal E}(J_3,\Delta_{1(3)})$ energy surfaces for different phases.  The upper panel, Fig.~\ref{Fig_PhDJ1J3}(a), is for the full $XXZ$ model, with the top edge corresponding to both $J_1$ and $J_3$ terms in the $XY$ limit and bottom edge to the fully isotropic Heisenberg model. Fig.~\ref{Fig_PhDJ1J3}(b) is for the partial $XXZ$ case, with $\Delta_3\!=\!1$ and  $\Delta_1$ varying from the Heisenberg (top edge) to the $XY$ limit of the $J_1$ term (bottom edge). Finally,  Fig.~\ref{Fig_PhDJ1J3}(c) closes the loop indicated in the inset of Fig.~\ref{Fig_PhDJ1J3}(b) by keeping $\Delta_1\!=\!0$ and interpolating $\Delta_3$ from the Heisenberg (top edge) to the $XY$ limit of the $J_3$ term (bottom edge). 

% ==============================================================================
% ==============================================================================

They are all constructed from the lines of the pairwise intersections of the $\Delta_{1(3)}$--$J_3$ energy surfaces for the FM, ZZ, dZZ, and Iz phases that are calculated using MAGSWT method described in the previous Sections. 

Although the full $XXZ$  phase diagram in Fig.~\ref{Fig_PhDJ1J3}(a) shows only slightly smaller region of the dZZ phase of very similar shape compared to that of the partial $XXZ$ case in Fig.~\ref{Fig_PhDJ1J3}(b), it is missing the Iz phase entirely, as discussed above. The additional cut along the $\Delta_3$ axis in Fig.~\ref{Fig_PhDJ1J3}(c), which interpolates between the full and partial $XXZ$ cases with the $J_1$ term fixed to the $XY$ limit, demonstrates a significant range occupied by the unexpected escapist Iz phase in this extended parameter space along the $J_3$ and $\Delta_{1(3)}$ directions. 

The classical FM-Sp and Sp-ZZ transition points $J_{3,c1}$ and $J_{3,c2}$ are marked by vertical dashed lines in  Fig.~\ref{Fig_PhDJ1J3}; one can see how dramatically the quantum phase boundaries deviate from those classical values.

Some of the presented phase diagrams, for the full and partial $XXZ$ models in Figs.~\ref{Fig_PhDJ1J3}(a) and \ref{Fig_PhDJ1J3}(b), have been compared with the DMRG results in Ref.~\cite{j1j3}, demonstrating close qualitative and quantitative agreements. The noncollinear Sp phase is completely wiped out from the phase diagrams in both MAGSWT and DMRG approaches, with the FM and ZZ phases expanding well beyond their classical boundaries. The regions of the dZZ phase in the full and partial $XXZ$ models are somewhat narrower in their $J_3$ extent in the DMRG results, while the area of the Iz phase in the partial $XXZ$ model is  in rather close accord between the two methods.

 Although DMRG  has shown a very narrow slice of the Iz phase between FM and ZZ  regions in the full $XXZ$ version of the model for $\Delta\!\alt\!0.35$~\cite{j1j3}, compared with the direct FM-ZZ transition by MAGSWT in Fig.~\ref{Fig_PhDJ1J3}(a), the energy analysis of  Sec.~\ref{Sec:J1J3resultsE} shows that these discrepancies  correspond to very small energy margins. They can  be ascribed to  the approximate nature of the MAGSWT approach and sensitivity of certain transitions to the higher-order corrections to it, and may be affected by the finite-size effects in DMRG as well. 
 
 With the overall topology of the phase diagrams and the identity of the competing phases captured correctly by MAGSWT, the provided results show rather remarkable agreement with the DMRG  data, underscoring the power of the MAGSWT method for exploring quantum phase diagrams in frustrated spin systems.

In summary, for the $J_1$--$J_3$ honeycomb-lattice model, the MAGSWT analysis provides a clear picture: quantum fluctuations strongly favor collinear phases, FM, ZZ, dZZ, and Iz, and disfavor the spiral, resulting in a quantum phase diagram that is qualitatively altered from the classical one and in close agreement with numerical findings. The success of this relatively simple analytical method in reproducing the numerical phase diagram lends support to the latter and offers additional insights into the nature of each phase, such as the magnitude of quantum contributions, the role of anisotropies, etc. 

We now turn to the $J_1$--$J_2$ model to test the MAGSWT approach in a more challenging situation, where the classical ground state harbors a macroscopic degeneracy.

%-------------------------------------------------------------------------------
\section{$J_1$--$J_2$ AF-AF Model}
\label{Sec:J1J2}
%-------------------------------------------------------------------------------

%-------------------------------------------------------------------------------
\subsection{Model and some background}
\label{Sec:J1J2model_history}
%-------------------------------------------------------------------------------

Here we follow the same narrative as in Sec.~\ref{Sec:J1J3}:  we briefly discuss some of the previous studies of the $J_1$--$J_2$ model,  show the classical and  LSWT results for relevant  phases, and follow with the demonstration of the MAGSWT outcomes for this model for $S\!=\!1/2$. 

As was mentioned above, the {\it classical} phase diagram of the mixed FM-AF $J_1$--$J_2$--$J_3$ honeycomb-lattice model, Ref.~\cite{Rastelli79}, can be mapped onto that of the AF $J_1$--$J_2$--$J_3$ model on the same lattice, with a proper relabeling of the phases. The degeneracies in the  classical phase diagram have also been mentioned in the early work~\cite{Rastelli79},  but the quantum ObD selection from the classically degenerate manifold have not been discussed until more recently~\cite{j1j2-arun}.  

The quantum $S\!=\!\frac12$ $J_1$--$J_2$--$J_3$ model has been searched for the exotic  states~\cite{Oitmaa91,j1j2j3-ed2001,Oitmaa11}, motivated by expectations of strong fluctuations due to low coordination number of the lattice.  The systematic studies have been devoted to the  $J_1$--$J_2$ AF model in the Heisenberg and $XY$ limits, originally suggesting spin-liquid states, but uncovering a number of unexpected ones, and also finding some phases that expand considerably from their classically prescribed regions~\cite{j1j2-rigol2,j1j2-bishop,j1j2-gong,j1j2-steve1,j1j2-steve2}. We will discuss some these prior results in more detail below.

The anisotropic $XXZ$  $J_1$--$J_2$ AF model on the honeycomb lattice is given by
\begin{eqnarray}
\label{eq_HJ1J2}
{\cal H}=\sum_{n=1,2}\sum_{\langle ij\rangle_n} J_n  \Big(S^{x}_i S^{x}_j+S^{y}_i S^{y}_j+\Delta S^{z}_i S^{z}_j\Big),
\end{eqnarray}
where the nearest-neighbor exchange, $J_1\!=\!1$, is an energy unit, and both exchanges are AF, $J_{1(2)}\!>\!0$. We consider only the easy-plane $XXZ$ anisotropy regime as before, $0\!\leq\!\Delta\!\leq\!1$, and focus on the case when it is the same in the $J_1$ and $J_2$  terms, $\Delta_{1}\!=\!\Delta_{2}\!=\!\Delta$, the ``full'' $XXZ$ case. 

%-------------------------------------------------------------------------------
\subsection{Classical phases and phase diagram}
\label{Sec:J1J2classical}
%-------------------------------------------------------------------------------

\begin{figure}[t]
\includegraphics[width=\linewidth]{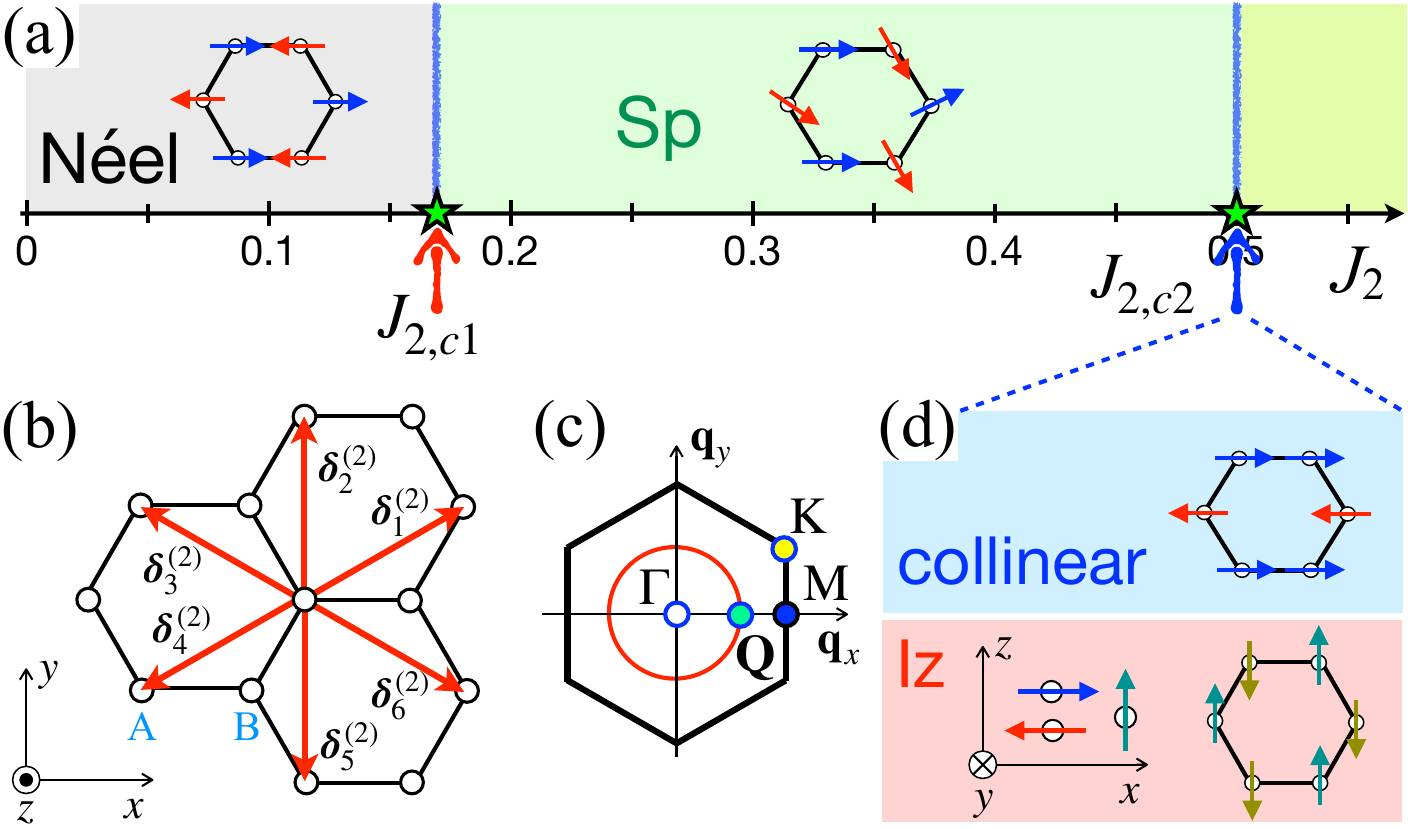}
\vskip -0.2cm
\caption{(a) Classical phase diagram of the $J_1$--$J_2$ model (\ref{eq_HJ1J2}) for any $0\!\leq\!\Delta\!\leq\!1$. Sketches of the N\'eel and Sp states illustrate their spin orders with blue and red arrows belonging to A and B sublattices. The transition points $J_{2,c1}\!=\!1/6$ and $J_{2,c2}\!=\! 1/2$ are indicated, see the text. (b) The  honeycomb lattice lattice with the second-neighbor vectors, ${\bm \delta}^{(2)}_\alpha$. (c) BZ of the honeycomb lattice with the high-symmetry $\Gamma$, M, and K points and the ${\bf Q}$ vector of the spiral selected by ObD from the degenerate states on the red contour. (d) Sketches of the collinear and Iz states. Axes show out-of-plane spin direction in the Iz phase and in-plane for the other phases.}
\label{Fig_1DJ1J2}
\vskip -0.3cm
\end{figure}

As in the case of the $J_1$--$J_3$ model in Sec.~\ref{Sec:J1J3}, all states minimizing the classical energy are coplanar with the $x$-$y$ plane, so that the classical phase diagram shown in Fig.~\ref{Fig_1DJ1J2}(a) is the same for any value of the easy-plane anisotropy $\Delta$~\cite{Rastelli79}. The phase diagram has the following phases: N\'eel phase in the regime of $J_2\!\leq\!J_{2,c1}\!=\!1/6$,  and a planar co-rotating spin spiral (Sp) for $J_2\!>\!1/6$, which is similar to the one in the $J_1$--$J_3$ model. 

In the N\'eel state, the 1D zigzag chains with the AF-ordered spins are arranged antiferromagnetically. Asymptotically, for $J_2\!\rightarrow\!\infty$ not shown in Fig.~\ref{Fig_1DJ1J2}(a), the model decouples into two triangular lattices  connected by the $J_2$ couplings, which are made of the sublattices A and B of the honeycomb lattice, as is shown in Fig.~\ref{Fig_1DJ1J2}(b), with each of them ordering into the 120$\degree$ state. 

The Sp phase, while similar to the one discussed in Sec.~\ref{Sec:J1J3}, has some important differences, making it substantially more complex. It borders the N\'eel state, but for $J_2\!>\!1/6$, its ordering vector ${\bf Q}$ belongs to a classically degenerate manifold of spirals that reside on a 1D contour in the momentum space, shown schematically in Fig.~\ref{Fig_1DJ1J2}(c), which evolves continuously with $J_2$. As was shown in Ref.~\cite{j1j2-arun} by an explicit calculation of the $O(S)$ energies for all states in such a manifold, the ObD effect selects the ordering vector ${\bf Q}$ that continuously migrates between the $\Gamma$ to M point for $J_2$ changing from $J_{2,c1}\!=\!1/6$ to $J_{2,c2}\!=\!1/2$, and from the M to K point for $J_2$ from $1/2$ to $\infty$, with these two different sectors of $J_2$ highlighted in Fig.~\ref{Fig_1DJ1J2}(a). 
 
% ==============================================================================
% ==============================================================================

With this ObD insight, the $J_{2,c2}\!=\!1/2$ point is special, as it corresponds to another commensurate and collinear spin state, in which the AF-ordered zigzag chains arrange ferromagnetically, see Fig.~\ref{Fig_1DJ1J2}(d). In that sense, the Sp phase is similar to the one in Sec.~\ref{Sec:J1J3}: it also continuously interpolates between the  state with the ordering vector at the $\Gamma$ point (N\'eel) and the state with ${\bf Q}$ at the M point (collinear), having them as  limiting cases~\cite{j1j2-arun}. 

As we will show below for the quantum case, the collinear state expands considerably from its classical range consisting of   just one $J_2$ value of 0.5. Its main competitor for the groundstate is the familiar Iz state, see Fig.~\ref{Fig_1DJ1J2}(d), and together they nearly eliminate the entire Sp phase from the $S\!=\!\frac12$ phase diagram of this model. 

In this work, we consider the $J_1$--$J_2$ $S\!=\!\frac12$ phase diagram and the role of quantum contributions in it only for the range of $J_2$ from 0 to 0.5. For the larger values of $J_2$, Ref.~\cite{j1j2-rigol2} shows that  the much-expanded collinear phase has a transition to another much-expanded state at $J_2\!\approx\!1.3$. It is the state in which the 120$\degree$ orders of the two sublattices from the  $J_2\!\rightarrow\!\infty$  limit are locked ferromagnetically, with no sign of the surviving Sp phase. 

Another aspect of the previous works will not be included here. In the Heisenberg limit of the  $J_1$--$J_2$ $S\!=\!\frac12$ model, instead of the spin-liquid states suspected earlier, DMRG and other methods~\cite{j1j2-steve1,j1j2-steve2,j1j2-bishop} have identified two {\it nonmagnetic} ordered valence-bond states (VBSs), which occupy the range from the expanded N\'eel boundary of  $J_{2}\!\approx\!0.22$ to $J_{2}\!\approx\!0.7$. In this work, we will  consider   competition of only {\it magnetically ordered} states. 

A simple algebra yields the classical energies of  the N\'eel, collinear, and Iz states within the model (\ref{eq_HJ1J2}),  per number of atomic unit cells of the honeycomb lattice $N_A$ and in units of $J_1$, as given by
\begin{align}
\label{eq_Ecl2}
&E_{cl}^{\rm N}=- 3S^2(1-2J_2),\ \ \  E_{cl}^{\rm Co}=-S^2(1+2J_2), \\
& E_{cl}^{\rm Iz}=- 3S^2\Delta (1-2J_2), \nonumber
\end{align}
where N\'eel and collinear states are independent of $\Delta$ as they are coplanar with the $x$-$y$ plane. These expressions are valid for any $J_2$, inside or outside the states' stability regions. Trivially, for $\Delta\!=\!1$,  N\'eel and Iz states are degenerate as the in-plane and out-of-plane N\'eel ordering is identical in the Heisenberg limit. 

With some algebra, using the single-${\bf Q}$ ansatz for the Sp state, its energy can  be obtained~\cite{Rastelli79} as given by, 
\begin{align}
\label{eq_Ecl_sp2}
E_{cl}^{\rm Sp}=-3S^2\left(\Re\left[e^{i\varphi}\gamma_{-\bf Q}\right]
-2J_2\gamma^{(2)}_{\bf Q}\right),
\end{align}
per $J_1N_A$. With the ObD insight for the choice of the ordering vector ${\bf Q}\!=\!(Q_x,0)$ along the $\Gamma$M line, $Q_x$ and $\varphi$ are defined as
\begin{eqnarray}
\label{eq_Qx2}
&&Q_x=\frac{2}{3}\cdot\arccos\left(\frac{1}{16J_2^2}-\frac54\right),\\
&&\varphi=-\frac{Q_x}{2}+\arctan\left(\frac{\sin (3Q_x/2)}{2+\cos (3Q_x/2)}\right), \nonumber
\end{eqnarray}
with all momenta in units of $1/a$. In Eq.~(\ref{eq_Ecl_sp2}), the nearest-neighbor hopping amplitude is defined in Eq.~(\ref{eq_gks}) and the second-neighbor hopping amplitude is 
\begin{align}
\label{eq_gks2}
\gamma^{(2)}_{\mathbf{q}}=\frac{1}{6}\sum_{\alpha} \cos \mathbf{q}{\bm \delta}^{(2)}_\alpha,
\end{align}
with the primitive vectors ${\bm \delta}^{(2)}_\alpha$ shown in Fig.~\ref{Fig_1DJ1J2}(b).
One can verify that the ordering vector ${\bf Q}$ in (\ref{eq_Qx2}) is continuously migrating as a function of $J_2$ from the $\Gamma$ point at $J_{2,c1}\!=\!1/6$ to the M point [$=\!(2\pi/3,0)$] at $J_{2,c2}\!=\!1/2$, in agreement with the discussion above. 

% ==============================================================================
\begin{figure}[t]
\includegraphics[width=\linewidth]{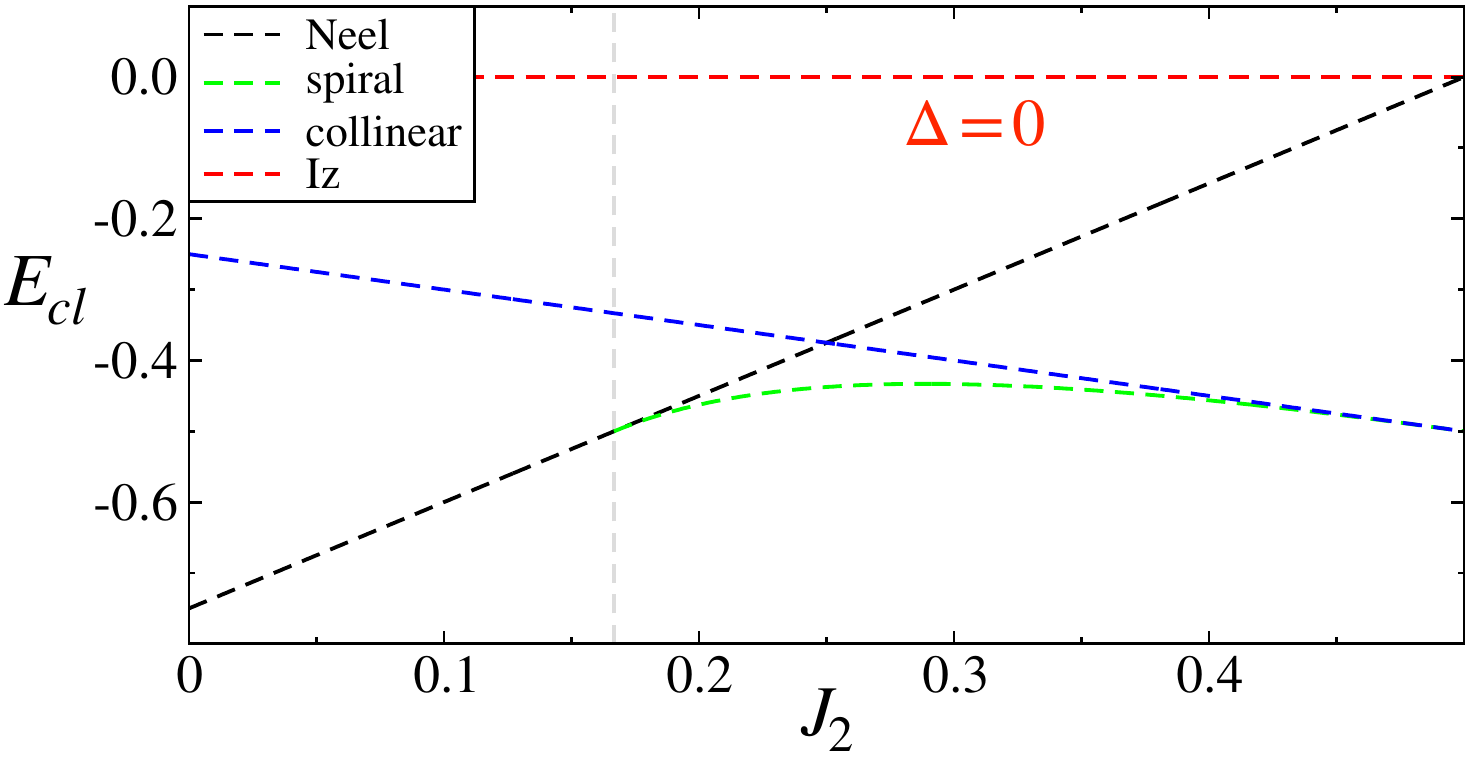}
\vskip -0.2cm
\caption{Classical energies of the N\'eel, Sp, collinear, and Iz states states as a function of $J_2$ from Eqs.~(\ref{eq_Ecl2}) and (\ref{eq_Ecl_sp2}) for $S\!=\!\frac12$. Vertical dashed line is the  N\'eel-Sp transitions, see Fig.~\ref{Fig_1DJ1J2}(a). For the Iz state, the limiting case $\Delta\!=\!0$ is shown; for $\Delta\!=\!1$ it is identical to N\'eel.}
\label{Fig_EclJ1J2}
\vskip -0.3cm
\end{figure}
% ==============================================================================

In Figure~\ref{Fig_EclJ1J2}, we show classical energies of all four states from Eqs.~(\ref{eq_Ecl2}) and (\ref{eq_Ecl_sp2}),  in the relevant range of $J_2$ and using spin $S\!=\!\frac12$. Vertical dashed line marks the N\'eel-Sp  transition, and the shown results for all states except for Iz are independent of the $XXZ$ anisotropies within the discussed easy-plane regime. The Sp state is continuously interpolating N\'eel and collinear states, as discussed above. For the Iz state,  the energy is shown in the $XY$ limiting case ($\Delta\!=\!0$), while in the isotropic $\Delta\!=\!1$ limit it is degenerate with the N\'eel state. 

In a significant similarity to the $J_1$--$J_3$ case discussed in Sec.~\ref{Sec:J1J3}, the Iz state is considerably higher than the rest of the states for $\Delta\!=\!0$, and the N\'eel and collinear states are akin to the FM and ZZ ones in the $J_1$--$J_3$ model. The differences are in the degeneracy of the Sp state and in that the collinear state is  classically stable only at $J_2\!=\!0.5$---the end-point of the $\Gamma$-M spiral phase.

In our analysis below, we will restrict attention to the magnetically ordered states: the N\'eel two-sublattice AF state with the ordering vector ${\bf Q}\!=\!0$, the collinear AF state with the four-site magnetic unit cell and ordering vector at the M point, the Iz state with the out-of-plane two-sublattice N\'eel order, and the spiral state described above. 

%-------------------------------------------------------------------------------
\subsection{LSWT}
\label{Sec:J1J2SWT}
%-------------------------------------------------------------------------------

Here we develop the standard LSWT  for all four classical states to pave the way for the MAGSWT.

The unit cell of the magnetic structure for all four states shown in Fig.~\ref{Fig_1DJ1J2}(a) and \ref{Fig_1DJ1J2}(d) is either that of the atomic unit cell of the honeycomb lattice from the start (N\'eel and Iz), or can be reduced to it using the staggered or rotated reference frames (collinear or Sp, respectively). Thus, the Hamiltonian matrix $\hat{\bf H}_{\bf q}$ in Eq.~(\ref{eq_LSWTmatrix}) is $4\times 4$  for all of them, with the LSWT matrices $\hat{\bf A}_{\bf q}$, $\hat{\bf B}_{\bf q}$ in Eq.~(\ref{eq_AB_2sub})  assuming the same form and the magnon eigenenergies given by the same general expression as in Eq.~(\ref{eq_E12_2sub}). While significant simplifications can be made for some of the states and in some $XXZ$ limits, guiding the search for the analytical form of the MAGSWT shifts of the chemical potential, this general formalism is still useful. 

%-------------------------------------------------------------------------------
\subsubsection{N\'eel}
\label{Sec:Neel}
%-------------------------------------------------------------------------------

In the N\'eel case,  the matrix elements of the LSWT matrices $\hat{\bf A}_{\bf q}$, $\hat{\bf B}_{\bf q}$ in Eq.~(\ref{eq_AB_2sub}) are given by
\begin{align}
A_{\bf q}&=3S\left(1-2J_2+J_2(1+\Delta)\gamma^{(2)}_{\bf q}\right), \nonumber\\
\label{eq_ABC_Neel}
B_{\bf q}&=\frac{3S}{2}(1-\Delta)\gamma_{\bf q}, \ \ C_{\bf q}=\frac{3S}{2}(1+\Delta)\gamma_{\bf q},\\
D_{\bf q}&=3S J_2(1-\Delta)\gamma^{(2)}_{\bf q}, \nonumber
\end{align}
with the hopping amplitudes given in Eqs.~(\ref{eq_gks}) and (\ref{eq_gks2}). As in all other cases considered here,  $A_{\bf q}$ and $D_{\bf q}$ are purely real, $B_{-\bf q}\!=\!B^*_{\bf q}$, and $C_{-\bf q}\!=\!C^*_{\bf q}$.

Formally, the diagonalization of the LSWT Hamiltonian can be approached differently in this case using the symmetry of the N\'eel state on the honeycomb lattice, which leads to the more symmetric structure of the $\hat{\bf H}_{\bf q}$ matrix. The approach consists of the unitary transformation that reduces the original $4\times 4$ matrix to the block-diagonal form of the two $2\times 2$ matrices, corresponding to the symmetric and antisymmetric combinations of the spin-flips. It is followed by the textbook Bogolyubov transformation for each  block~\cite{kopietz,maksimov16,maksimov22,Smit20}, leading to the analytical form of the magnon eigenenergies 
\begin{align}
\label{eq_E12_Neel_general}
\varepsilon_{\nu,{\bf q}}&=3S\sqrt{\widetilde{A}_{\nu,\bf q}^2-\widetilde{B}_{\nu,\bf q}^2}\, ,
\end{align}
with $\widetilde{A}_{\nu,\bf q}\!=\!A_{\bf q}+(-1)^\nu|B_{\bf q}|$, and  $\widetilde{B}_{\nu,\bf q}\!=\!D_{\bf q}+(-1)^\nu|C_{\bf q}|$.

There are further significant simplifications available in the Heisenberg and $XY$ limits of the model, which are useful for the subsequent MAGSWT insights. For the former, $B_{\bf q}\!=\!D_{\bf q}\!=\!0$, leading to the two degenerate branches
\begin{align}
\label{eq_E12_Heis_Neel}
\varepsilon_{1(2),{\bf q}}=&3S\sqrt{\Big(1-2J_2+2J_2\gamma^{(2)}_{\bf q}+\left|{\gamma}_{\bf q}\right|\Big)}\\
&\times\sqrt{\Big(1-2J_2+2J_2\gamma^{(2)}_{\bf q}-\left|{\gamma}_{\bf q}\right|\Big)},\nonumber
\end{align}
and for the latter, $B_{\bf q}\!=\!C_{\bf q}$,
\begin{align}
\label{eq_E12_XYNeel}
\varepsilon_{\nu,{\bf q}}=&3S\sqrt{1-2J_2}\\ 
&\times\sqrt{\Big(1-2J_2+2J_2\gamma^{(2)}_{\bf q}+(-1)^\nu\left|{\gamma}_{\bf q}\right|\Big)},\nonumber
\end{align}
with the second bracket in the lower magnon branch containing the ``offending'' element, which is responsible for the softening of the spectrum for $J_{2}\!>\!J_{2,c1}$ in both limits. 

%-------------------------------------------------------------------------------
\subsubsection{Iz}
\label{Sec:Iz2}
%-------------------------------------------------------------------------------

Since the Iz state is, essentially, an out-of plane N\'eel state, identical to it in the Heisenberg limit, one can use the transformations mentioned above and obtain expression for the two degenerate magnon branches as
\begin{align}
\label{eq_E12_Iz_general}
\varepsilon_{1(2),{\bf q}}&=3S\sqrt{\widetilde{A}_{z,\bf q}^2-\widetilde{B}_{z,\bf q}^2}\, ,
\end{align}
with $\widetilde{A}_{z,\bf q}\!=\!\Delta(1-2J_2)+2J_2\gamma^{(2)}_{\bf q}$, and  $\widetilde{B}_{z,\bf q}\!=\!\left|{\gamma}_{\bf q}\right|$, which simplify to 
\begin{align}
\label{eq_E12_Iz_2}
\varepsilon_{1(2),{\bf q}}=&3S\sqrt{\Big(\Delta(1-2J_2)+2J_2\gamma^{(2)}_{\bf q}+\left|{\gamma}_{\bf q}\right|\Big)}\\
&\times\sqrt{\Big(\Delta(1-2J_2)+2J_2\gamma^{(2)}_{\bf q}-\left|{\gamma}_{\bf q}\right|\Big)},\nonumber
\end{align}
with a similar structure in the second bracket as in (\ref{eq_E12_XYNeel}). 

%-------------------------------------------------------------------------------
\subsubsection{Collinear}
\label{Sec:Collinear}
%-------------------------------------------------------------------------------

In the collinear case,  the matrix  elements are 
\begin{align}
A_{\bf q}&=S\Big(1+J_2\big(2+(1+\Delta)\gamma^{(2)}_{2,\bf q}+(1-\Delta)\gamma^{(2)}_{13,\bf q}\big)\Big), \nonumber\\
B_{\bf q}&=\frac{S}{2}\Big((1+\Delta)\gamma_{1,\bf q} +(1-\Delta)\gamma_{23,\bf q}\Big),\nonumber\\
C_{\bf q}&=\frac{S}{2}\Big((1-\Delta)\gamma_{1,\bf q} +(1+\Delta)\gamma_{23,\bf q}\Big),\nonumber\\
\label{eq_ABCD_Co}
D_{\bf q}&=SJ_2\Big((1-\Delta)\gamma^{(2)}_{2,\bf q}+(1+\Delta)\gamma^{(2)}_{13,\bf q}\Big),
\end{align}
with the nearest-neighbor hopping amplitudes as in Eq.~(\ref{eq_gks_dZZ})
\begin{align}
\label{eq_gks_Co1}
\gamma_{1,\bf q}&=e^{i{\bf q}{\bm \delta}_1},\ \
\gamma_{23,\bf q}=e^{i{\bf q}{\bm \delta}_2}+e^{i{\bf q}{\bm \delta}_3}, 
\end{align}
 and the second-nearest-neighbor amplitudes given by
\begin{align}
\label{eq_gks_Co2}
\gamma^{(2)}_{2,\bf q}&=\cos {\bf q}{\bm \delta}^{(2)}_2,\ \
\gamma^{(2)}_{13,\bf q}=\cos {\bf q}{\bm \delta}^{(2)}_1+\cos {\bf q}{\bm \delta}^{(2)}_3,
\end{align}
with the ${\bm \delta}^{(2)}_\alpha$ vectors shown in Fig.~\ref{Fig_1DJ1J2}(b).

In this case,  the  $XY$ limit is instructive, leading to
\begin{align}
\label{eq_E12_Co}
\varepsilon_{\nu,{\bf q}}=&S\sqrt{1+2J_2}\\ 
&\times\sqrt{\left(1+2J_2+6J_2\gamma^{(2)}_{\bf q}+3(-1)^\nu \left|{\gamma}_{\bf q}\right|\right)},\nonumber
\end{align}
containing the same structural elements in the second bracket as in the previous cases, suggesting a common hint for the  MAGSWT shifts of the chemical potential.

%-------------------------------------------------------------------------------
\subsubsection{Sp}
\label{Sec:Sp2}
%-------------------------------------------------------------------------------
Finally, for the Sp state,  the matrix  elements are 
\begin{align}
A_{\bf q}&=3S\bigg(\Re\Big[e^{i\varphi}\gamma_{-\bf Q}\Big]-2J_2\gamma^{(2)}_{\bf Q} \nonumber\\
&\  \quad\quad+J_2\Big(\Delta\gamma^{(2)}_{\bf q}
+\frac{1}{2}\big(\gamma^{(2)}_{{\bf q}-{\bf Q}}+\gamma^{(2)}_{{\bf q}+{\bf Q}}\big)\Big)\bigg),\nonumber\\
\label{eq_ABC_SpJ1J2}
B_{\bf q}&=-\frac{3S}{2}\Big(\Delta\gamma_{\bf q}-
\frac{1}{2}\big(e^{i\varphi}\gamma_{{\bf q}-\bf Q}+e^{-i\varphi}\gamma_{{\bf q}+\bf Q}\big)\Big),\\
C_{\bf q}&=\frac{3S}{2}\Big(\Delta\gamma_{\bf q}+
\frac{1}{2}\big(e^{i\varphi}\gamma_{{\bf q}-\bf Q}+e^{-i\varphi}\gamma_{{\bf q}+\bf Q}\big)\Big),\nonumber\\
D_{\bf q}&=-3SJ_2\Big(\Delta\gamma^{(2)}_{\bf q}
-\frac{1}{2}\big(\gamma^{(2)}_{{\bf q}-{\bf Q}}+\gamma^{(2)}_{{\bf q}+{\bf Q}}\big)\Big),\nonumber
\end{align}
where ${\bf Q}$ and $\varphi$ are given in (\ref{eq_Qx2}) and we generalized results of Ref.~\cite{Rastelli79}, which considered only  $XY$ and Heisenberg limits. The magnon energies follow from (\ref{eq_E12_2sub}).

%-------------------------------------------------------------------------------
\subsection{Finding $\mu$}
\label{Sec:J1J2mu}
%-------------------------------------------------------------------------------

%%-------------------------------------------------------------------------------------------
\begin{figure}[t]
\includegraphics[width=\linewidth]{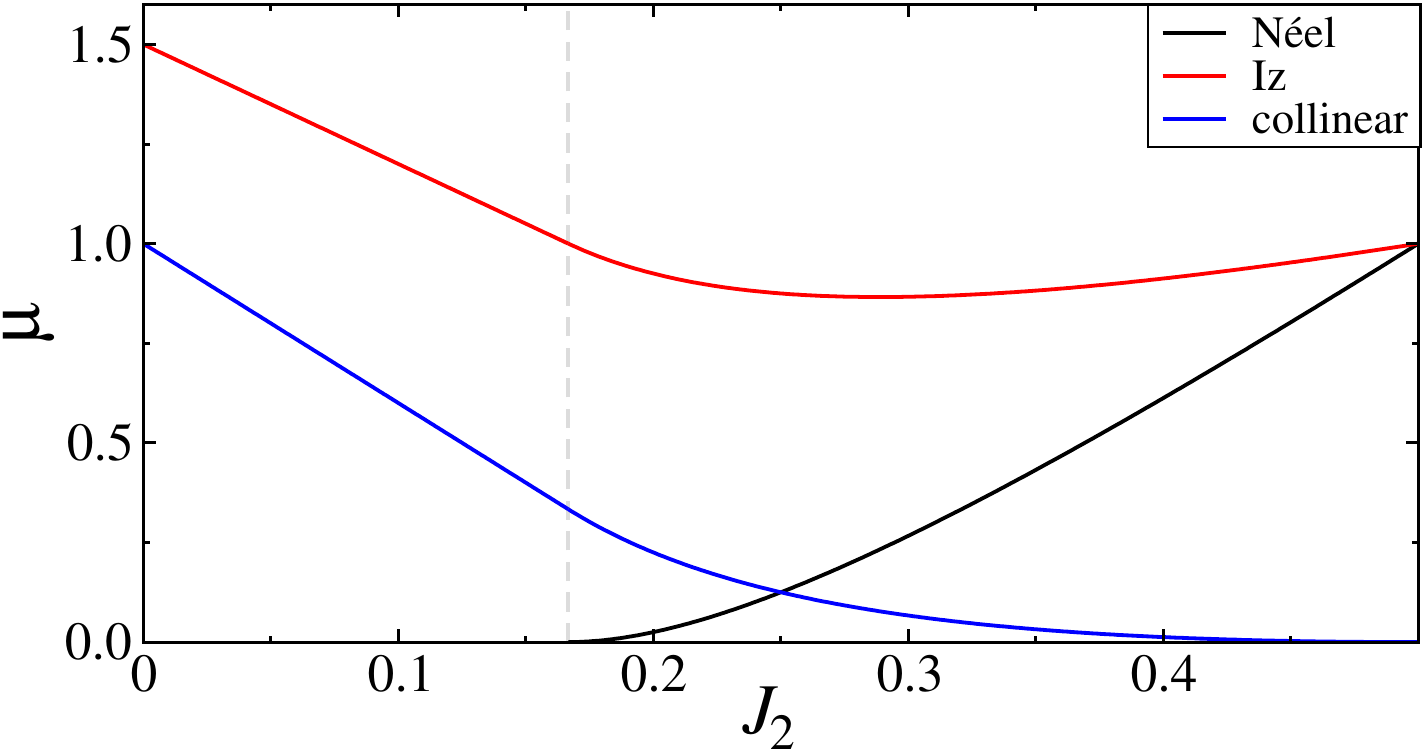}
\vskip -0.2cm
\caption{The minimal $\mu$ for the N\'eel, Iz,  and collinear  states, Eqs.~(\ref{eq_mu_Neel}), (\ref{eq_mu_Iz}),  and (\ref{eq_mu_Co}) respectively; for the Iz state $\Delta\!=\!0$.}
\label{fig:mu_vs_J2}
\vskip -0.4cm
\end{figure}
%%-------------------------------------------------------------------------------------------

Given the practice of Sec.~\ref{Sec:J1J3mu}, we follow the same strategy for finding the minimal MAGSWT shift of the chemical potential $\mu$ by examining the analytical expressions for the magnon bands obtained above in the limiting $XY$  and Heisenberg cases. It appears that for all three states in question, N\'eel, Iz, and collinear, the structure of the offending part of the magnon energy contains the same combination, $\widetilde{\gamma}_{\bf q}\!=\!\left|{\gamma}_{\bf q}\right|-2J_2\gamma^{(2)}_{\bf q}$, offset by different ${\bf q}$-independent terms. This suggests that in all three cases,
\begin{align}
\label{eq_musJ1J2}
\mu = a + b\widetilde{\gamma}_{{\bf Q}_{\rm max}}, 
\end{align}
where ${\bf Q}_{\rm max}$ is the momentum at which $\widetilde{\gamma}_{\bf q}$ achieves maximal value for a given $J_2$. As in Sec.~\ref{Sec:J1J3mu}, this condition is equivalent to the search of the ordering vector associated with the classical energy minimum of the $J_1$--$J_2$ model (\ref{eq_HJ1J2}). The difference in the present case is that for the classical Sp region of $J_2$ this vector belongs to a contour of degenerate states; see Fig.~\ref{Fig_1DJ1J2}(c). 

Technically, any ${\bf Q}_{\rm max}$ from that contour is sufficient, because the augmented spectrum is to be built on the LSWT results. Below, we simply list explicit piecewise expressions for the shift of the chemical potential. One can verify that they resolve the problem of the stability of the LSWT spectra for all three states and for all values of $\Delta$.

With some straightforward algebra one can eliminate the notion of ${\bf Q}_{\rm max}$ from $\mu$ and have an explicit   expression for it in terms of $J_2$. For the N\'eel state, 
\begin{align}
\label{eq_mu_Neel}
\mu&=
\left\{\begin{array}{ll} 
0, &    J_2 \leq  1/6, \\
S\,\displaystyle{\frac{(6J_2-1)^2}{4J_2}}, &  1/6 < J_2 \leq 1/2.
\end{array} \right.
\end{align}

Similarly, for the Iz state,
\begin{align}
\label{eq_mu_Iz}
\mu-\mu_0&=
\left\{\begin{array}{ll} 
0, &    J_2 \leq  1/6, \\
S\,\displaystyle{\frac{(6J_2-1)^2}{4J_2}}, &  1/6 < J_2 \leq 1/2.
\end{array} \right.
\end{align}
with the additional offset $\mu_0\!=\!3S(1-\Delta)(1-2J_2)$.

For the collinear state,
\begin{align}
\label{eq_mu_Co}
\mu&=
\left\{\begin{array}{ll} 
2S(1-4J_2), &    J_2 \leq  1/6, \\
S\,\displaystyle{\frac{(1-2J_2)^2}{4J_2}}, &  1/6 < J_2 \leq 1/2.
\end{array} \right.
\end{align}
As in the case of the $J_1$--$J_3$ model, the resulting $\mu$ for the N\'eel and collinear states are independent of the $XXZ$ parameters $\Delta$, and the Iz state depends on it via a simple shift.  Because of this shift, the MAGSWT spectrum in the Iz case, and its quantum energy contribution (\ref{eq_dE}) derived from it, are fully independent of the anisotropy parameters $\Delta$, also the same as in Sec.~\ref{Sec:J1J3mu}.

Altogether, for a given $J_2$,  Eqs.~(\ref{eq_mu_Neel}), (\ref{eq_mu_Iz}),  and (\ref{eq_mu_Co}) define $\mu$ for the three states. In Figure~\ref{fig:mu_vs_J2}, we present their plot for $S\!=\!\frac12$; for the Iz state $\Delta\!=\!0$ is chosen. Sp state is classically stable throughout its range of existence and does not need MAGSWT augmentation. 

%-------------------------------------------------------------------------------
\subsection{Results, energies}
\label{Sec:J1J2resultsE}
%-------------------------------------------------------------------------------

Comparison of the $O(S)$ energies (\ref{eq_E}) for the competing magnetically ordered states throughout the parameter space of the model (\ref{eq_HJ1J2}) can now be readily performed.

\begin{figure}[t]
\includegraphics[width=\linewidth]{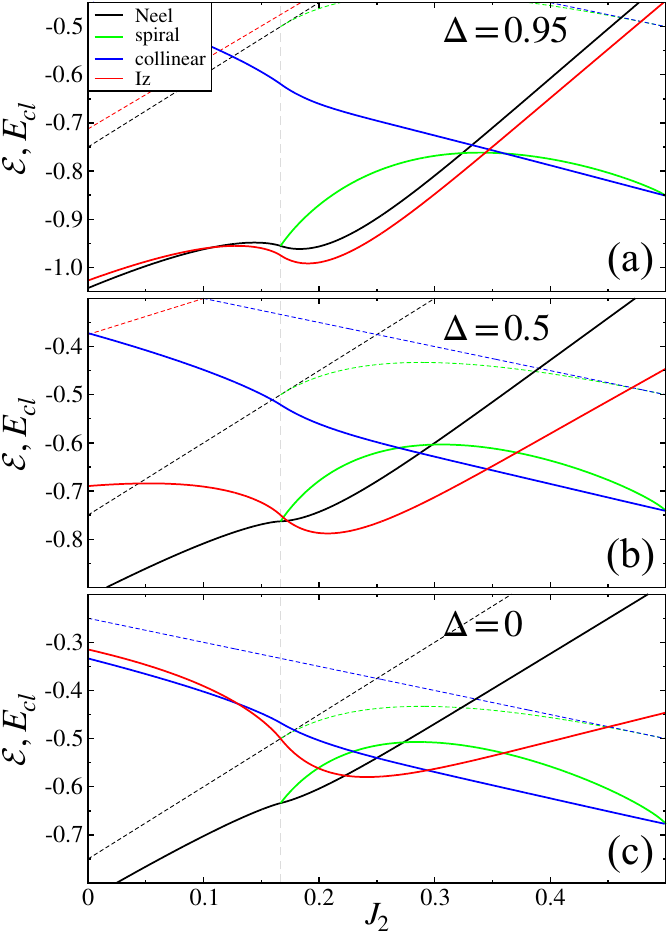}
\vskip -0.2cm
\caption{Energies of the N\'eel, Sp, Iz, and collinear states for $S\!=\!1/2$ vs $J_2$. Dashed lines are the classical energies, Eqs.~(\ref{eq_Ecl2}) and (\ref{eq_Ecl_sp2}), and solid lines are ${\cal E}\!=\!E_{cl}+\delta E$ from (\ref{eq_E}). Vertical dashed line is the classical  N\'eel-Sp boundary from Fig.~\ref{Fig_1DJ1J2}. (a) $\Delta\!=\!0.95$, (b) $\Delta\!=\!0.5$, and (c)  $\Delta\!=\!0$ ($XY$ limit).} 
\label{Fig_E_J2_all}
\vskip -0.4cm
\end{figure}

Figure~\ref{Fig_E_J2_all} shows representative results that illustrate such a competition along  $J_2$-cuts through the phase diagram for three  choices of the $XXZ$ parameter $\Delta\!=$0.95, 0.5, and 0.  We offset $\Delta$ from the Heisenberg limit in Fig.~\ref{Fig_E_J2_all}(a) because Iz and N\'eel states are degenerate in it. The dashed lines are the  classical energies from Fig.~\ref{Fig_EclJ1J2} and  Eqs.~(\ref{eq_Ecl2}) and (\ref{eq_Ecl_sp2}), and solid lines are the ${\cal E}$ energies obtained using Eq.~(\ref{eq_dE}) for $S\!=\!1/2$.  Vertical dashed line is the classical N\'eel-Sp boundary, $J_{2,c1}$, from Fig.~\ref{Fig_1DJ1J2}(a). 

While there are significant similarities with the energy comparison of the states in the $J_1$--$J_3$ model in Figs.~\ref{Fig_Es_partial} and~\ref{Fig_Es_full}, such as an upward arcs of the Sp energy, losing to the neighboring phases and underscoring, once more,  its lack of competitiveness, there are several differences that are worth highlighting. 

All competing states in Fig.~\ref{Fig_E_J2_all} are strongly fluctuating, with the Iz phase being competitive for all  $XXZ$ anisotropies, including  proximity to the Heisenberg limit, which is not the case in the $J_1$--$J_3$ model. In the same limit, the Sp state is also able to survive as a ground state in a narrow region of $J_2$. Although some of these features are going to be ``hidden'' in the true quantum phase diagram by the nonmagnetic VBS states that are not considered here, the competition of the N\'eel and Iz phase in this regime may require more detailed  study with the unbiased numerical methods. 

The N\'eel and collinear phases are expanding significantly from their classical ranges in all  panels of Fig.~\ref{Fig_E_J2_all}, exterminating  Sp state from most of the parameter space, all in agreement with  previous works~\cite{j1j2-rigol2,j1j2-bishop,j1j2-gong,j1j2-steve1,j1j2-steve2}. However, it is the Iz state that stays out remarkably in the anisotropic cases, Fig.~\ref{Fig_E_J2_all}(b) and~\ref{Fig_E_J2_all}(c).

Iz state fully confirms its reputation of a daring escapist state, with the downward energy renormalization exceeding that of any competing state threefold. In Fig.~\ref{Fig_E_J2_all}(c),  in the $XY$ limit of the model (\ref{eq_HJ1J2}), the quantum contribution $\delta E$ (\ref{eq_dE}) constitutes its {\it entire} energy. While its groundstate range narrows somewhat compared to Fig.~\ref{Fig_E_J2_all}(b), this may be related to the approximations of the MAGSWT, as the higher-order terms can further contribute  to the competition.

We reiterate here that the original discovery of the Iz  N\'eel-like state, with the ordered moments along the $z$ axis despite the model (\ref{eq_HJ1J2}) having no out-of-plane $S^zS^z$ interactions in its $XY$ limit, was totally unexpected~\cite{j1j2-steve2}. While, ideologically, a rationalization for its appearance as due to potentially large quantum fluctuations has been made, our present study offers the first explicit demonstration of the viability of such a scenario from the most natural perspective of the magnetically ordered state. 

We would also like to emphasize, once again,  that MAGSWT enables an easy access to quantitative insights and detailed  analysis of the energy competition of the classically unstable states.

% ==============================================================================
% ==============================================================================

%-------------------------------------------------------------------------------
\subsection{Results, phase diagram}
\label{Sec:J1J2resultsPhD}
%-------------------------------------------------------------------------------

% ==============================================================================
\begin{figure}[t]
\includegraphics[width=\linewidth]{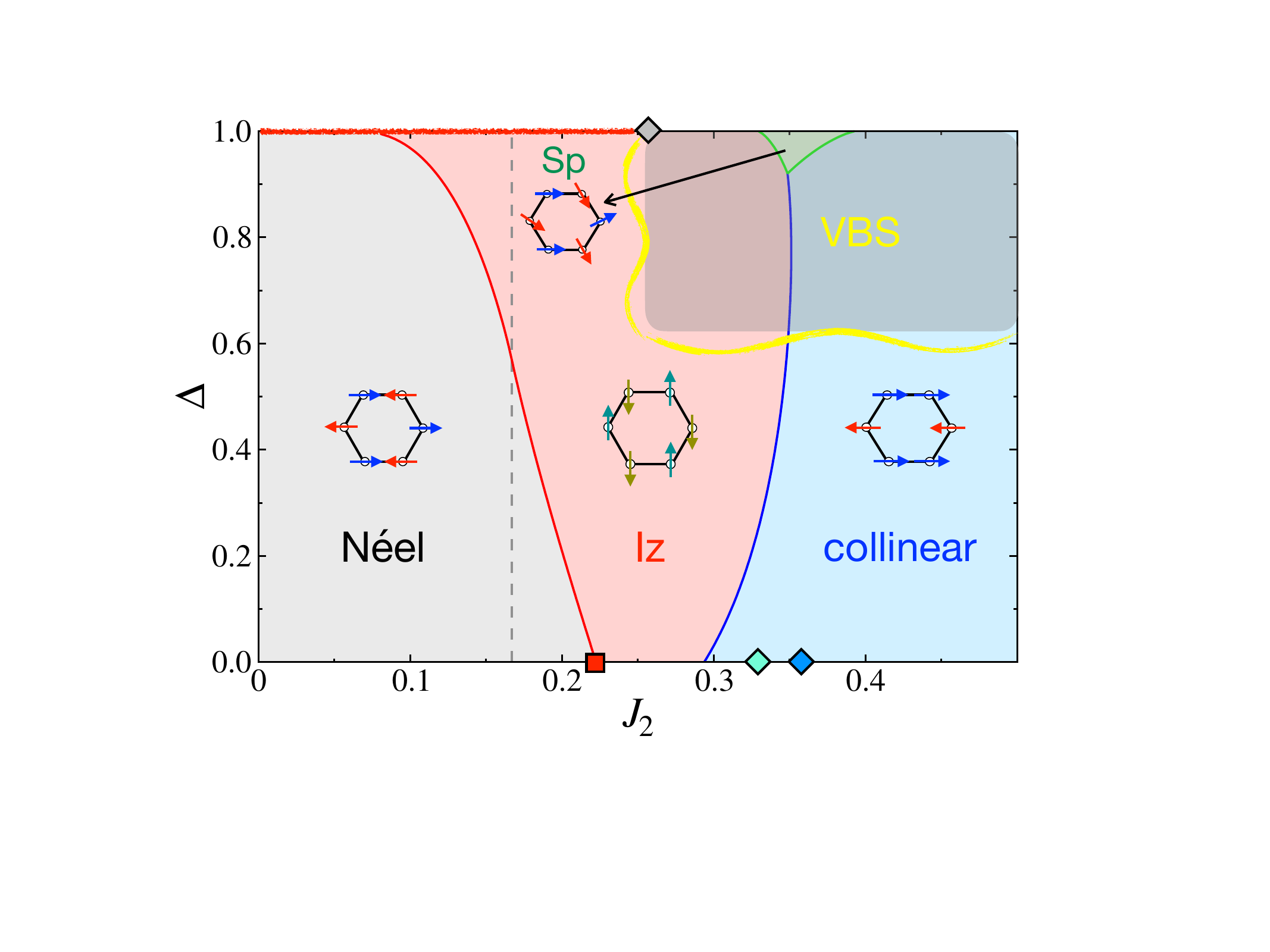}
\vskip -0.2cm
\caption{Phase diagram of the $J_1$--$J_2$ $XXZ$ model (\ref{eq_HJ1J2}) for $S\!=\!1/2$,  obtained by MAGSWT for the magnetically ordered states, with phases and their sketches identified. Vertical dashed line is the classical N\'eel-Sp phase boundary from Fig.~\ref{Fig_1DJ1J2}(a). Symbols are DMRG phase boundaries~\cite{j1j2-steve2,j1j2-steve1,Sedrakyan22}, see the text.  Horizontal line indicates that Iz and N\'eel states are degenerate at $\Delta\!=\!1$. Shaded region with curved borders indicates the approximate range dominated by the nonmagnetic VBS  states, as suggested by numerical results.}
\label{Fig_PhD_J2}
\vskip -0.4cm
\end{figure}
% ==============================================================================

We conclude with the MAGSWT phase diagram of the $S\!=\!1/2$ $J_1$--$J_2$ $XXZ$ model  (\ref{eq_HJ1J2}) for the magnetically ordered phases in the $J_2$--$\Delta$ plane, shown in Figure~\ref{Fig_PhD_J2}. The phase boundaries are drawn from the pairwise intersections of the ${\cal E}(J_2,\Delta)$ energy surfaces for the four competing states considered above. As is the case for the $J_1$--$J_3$ model in Sec.~\ref{Sec:J1J3resultsPhD}, the computational ease of finding the $O(S)$ MAGSWT energies in the full parameter space of the model is rather remarkable.

The main results are on full display: both N\'eel and collinear states expand well beyond their original regions (a single $J_2 = 0.5$ value in the latter case); a large swath of the phase diagram is occupied by the escapist Iz state, and---apart from a small region near the Heisenberg limit---the originally dominant Sp phase is nearly squeezed out. In these broad strokes, and in close agreement with numerical findings~\cite{j1j2-steve2,j1j2-steve1,j1j2-bishop}, the quantum phase diagram is qualitatively altered from the classical one in Fig.~\ref{Fig_1DJ1J2}(a), with its sole N\'eel-Sp classical boundary shown by the faint vertical dashed line, and is dominated by collinear magnetic orders.

With the broad conclusions in close agreement, there are significant qualitative and quantitative differences from  previous results that call for further investigation. Since we only consider magnetically ordered states, the nonmagnetic VBS phases are not accounted for in this study, with the approximate range dominated by them~\cite{j1j2-steve2,j1j2-steve1,j1j2-bishop} sketched  in Fig.~\ref{Fig_PhD_J2}. Not only does this region eliminate the remnants of the Sp phase, but it also carves out significant portions from the collinear and Iz phases in the vicinity of the Heisenberg limit. 

A quantitative difference of the phase diagram in Fig.~\ref{Fig_PhD_J2} from the previous studies~\cite{j1j2-steve2,j1j2-steve1,j1j2-bishop} is the behavior of the N\'eel-Iz boundary. While in the $XY$ limit the MAGSWT results are nearly coincident with the ones from the earlier DMRG study, which discovered the Iz phase, $J_2\!\approx\!0.22$ in DMRG (marked by the red square in  Fig.~\ref{Fig_PhD_J2}) against $J_2\!\approx\!0.222$ in MAGSWT, the $\Delta$-dependence of this boundary is different from the later study based on a coupled cluster method~\cite{j1j2-bishop}, which shows nearly constant-$J_2$ boundary. However, we note that the results of this later work also disagree quantitatively with the DMRG for the Heisenberg limit~\cite{j1j2-steve1}.  On the other hand, our analysis in Sec.~\ref{Sec:J1J2resultsE} demonstrates that the out-of-plane N\'eel (Iz state) consistently outpaces the in-plane N\'eel state in the fluctuating part of the energy, making the shift of their phase boundary toward the lower values of $J_2$ quite natural, and calling for more numerically unbiased  studies of this boundary in the intermediate range of $\Delta$. 

It must also be noted that the two states, N\'eel and Iz, are indistinguishable in the Heisenberg limit, as is emphasized by the red line along the $\Delta\!=\!1$ axis in Fig.~\ref{Fig_PhD_J2}, and they may be difficult to distinguish numerically near that limit. For that same reason, the N\'eel-VBS transition in DMRG (marked by the gray diamond in  Fig.~\ref{Fig_PhD_J2}), is also an Iz-VBS one. 

For the Iz-collinear boundary in the $XY$ limit in Fig.~\ref{Fig_PhD_J2}, the agreement is somewhat less close: $J_2\!\approx\!0.36$ in the earlier DMRG work~\cite{j1j2-steve2} (blue diamond) and  $J_2\!\approx\!0.33$ in the recent one~\cite{Sedrakyan22} (cyan diamond), against $J_2\!\approx\!0.294$ in MAGSWT. A justification of this numerical discrepancy may be that the Iz phase is a victim of its own strong downward renormalization in the $XY$ limit, which suggests that the role of the higher-order corrections to the MAGSWT energies is not negligible, also making the agreement for the N\'eel-Iz boundary fortuitous. The Iz–collinear boundary has a somewhat different structure in Ref.~\cite{j1j2-bishop}, which suggests a proliferation of the VBS state all the way to $\Delta\!=\!0$.  An additional DMRG study of this aspect would  be helpful. 

With the approximate nature of the phase diagram in  Fig.~\ref{Fig_PhD_J2} thoroughly exposed, one should not lose the perspective of its successes: the MAGSWT approach is able to correctly select the likely magnetic orderings that compete, and these coincide with those found by  computational methods. The method also successfully confirms the strong presence of the elusive Iz phase in the anisotropic $J_1$--$J_2$ model.

In summary, for the $J_1$--$J_2$ model, MAGSWT provides evidence of the unexpected quantum ground states among the magnetic orders, with the spiral states consistently higher in energy. These findings align qualitatively with the existing numerical work, reinforcing the conclusion that strong frustration plus quantum fluctuations lead either to different collinear orders or to nonmagnetic VBS phases, and call for more studies of this model. The success of this relatively simple semi-analytical method, once again, offers additional insights into the nature of the magnetic phases and their competition. 

%-------------------------------------------------------------------------------
\section{Conclusions}
\label{Sec:Conclusions}
%-------------------------------------------------------------------------------

We have demonstrated that the minimally-augmented spin-wave theory (MAGSWT) provides a powerful and efficient means to explore the phase diagrams of  quantum magnets. By introducing a positive shift of the magnon chemical potential, MAGSWT extends the $1/S$ expansion for the magnetically ordered states beyond classical stability limits and yields quantum groundstate energies that serve as upper bounds for those states to order $O(S)$. 

This approach enabled us to construct the faithful phase diagrams of two paradigmatic honeycomb-lattice models in the quantum $S\!=\!1/2$ limit---the $J_1$--$J_3$ FM-AF and $J_1$--$J_2$ AF models---for the collinear quantum phases that replace or extend the classical ones. Our results are in good qualitative and semi-quantitative agreement with state-of-the-art numerical studies for these models, correctly capturing the emergence of unexpected quantum phases and the suppression of classically favored spiral orders by quantum fluctuations.

In the $J_1$--$J_3$ model, quantum fluctuations stabilize the double-zigzag and out-of-plane Iz phases between the FM and zigzag orders, wiping out the intermediate spiral phase that is present classically, and provide a substantial expansion of the FM and zigzag  phases compared to the classical picture. These findings lend support to earlier numerical results that reported the same set of quantum ground states and a close agreement in the locations of phase transitions. 

For the $S\!=\!1/2$ $J_1$--$J_2$ model, MAGSWT shows that quantum fluctuations favor the N\'eel, collinear AF, and Iz orders over the degenerate spiral manifold of states. In particular, the Iz phase was found to occupy a large portion of the phase diagram, consistent with earlier DMRG findings of an unexpected Ising order in the $XY$ limit of this model. Here, the success of MAGSWT is in giving an explicit demonstration of the viability of large quantum fluctuations as the stabilization mechanism for this escapist state from the natural perspective of magnetically ordered states, even when such a state is classically unstable anywhere in the phase diagram.

Overall, our study showcases the utility of MAGSWT as a relatively simple semi-analytical approach to the phase diagrams of quantum magnets. It complements numerical methods by providing physical insight into which states are competitive and by yielding approximate phase boundaries that agree well with much more numerically intensive calculations. 

Looking forward, there are several potential directions to extend the minimally-augmented spin-wave approach. One challenge is to generalize MAGSWT to states that are not classical extrema, i.e., those that would produce linear terms in a spin-wave expansion. This includes generic noncollinear states or states in models with spin-orbit-induced anisotropic exchanges, which necessarily induce off-diagonal couplings of the spin components. Developing a scheme to systematically handle those cases would broaden the applicability of the method to a wider class of magnets with complex couplings.  In this respect, a very recent development in Ref.~\cite{PavelKitaevFM25} is very promising.

Another prospective extension is to apply similar augmentation ideas to the other bosonic theories, such as the ones for nematic orders and non-magnetic VBS phases. In the case of the bond-operator theories one can envision introducing a parameter analogous to the shift of the chemical potential to enforce stability of a candidate state outside its mean-field stability region. This could allow the study of quantum phase transitions between VBS and magnetic phases on equal footing.

In conclusion, the minimally-augmented spin-wave theory offers a compelling addition to the toolkit for quantum magnetism. It requires modest computational effort, builds on well-understood SWT, yet  yields important insights into the energetics of the competing states and quantitative results that closely mirror those from large-scale numerical simulations.  

We anticipate that this approach will be equally useful in other contexts, such as multi-spin exchange models or field-induced phenomena, where finding the correct ground state is often nontrivial. Our study reinforces the perspective that many seemingly mysterious quantum phases can be understood as natural extensions of the classical states into the quantum regime with fluctuations properly accounted for. The MAGSWT framework makes this extension systematic and sheds light on the rich phase diagrams of these systems.

%=========================================================
\begin{acknowledgments}
Useful conversations with Pavel Maksimov, Mike Zhitomirsky, Fr\'ed\'eric Mila, and  Steven White are gratefully acknowledged.

This work was supported by the U.S. Department of Energy, Office of Science, Basic Energy Sciences under Award No.~DE-SC0021221. We would like to thank the Aspen Center for Physics and the Kavli Institute for Theoretical Physics (KITP) for hospitality during different stages of this project. The Aspen Center for Physics is supported by National Science Foundation Grant No.~PHY-2210452, and KITP is supported by the National Science Foundation under Grant No.~NSF PHY-2309135.
\end{acknowledgments}
%=========================================================

%=========================================================
\bibliography{refs_MAGSWT}
%==================================================================
\end{document}